# Bioconjugated oligonucleotides: recent developments and therapeutic applications

*Sebastien Benizri,[§†‡] Arnaud Gissot,[§†‡] Andrew Martin,[#] Brune Vialet,[§†‡] Mark W. Grinstaff,[*#] and Philippe Barthélémy*[§†‡]*

[§]Inserm U1212, F-33076 Bordeaux, France, [†]CNRS 5320, F-33076 Bordeaux, France, [‡]Université de Bordeaux, 146 rue Léo Saignat, F-33076 Bordeaux Cedex, France, [#]Departments of Biomedical Engineering, Chemistry, and Medicine, Boston University, Boston, 02215, Massachusetts, USA.

**ABSTRACT:** Oligonucleotide-based agents have the potential to treat or cure almost any disease, and are one of the key therapeutic drug classes of the future. Bioconjugated oligonucleotides, a subset of this class, are emerging from basic research and being successfully translated to the clinic. In this review, we first briefly describe two approaches for inhibiting specific genes using oligonucleotides - antisense DNA (ASO) and RNA interference (RNAi) – followed by a discussion on delivery to cells. We then summarize and analyze recent developments in bioconjugated oligonucleotides including those possessing GalNAc, cell penetrating peptides, α-tocopherol, aptamers, antibodies, cholesterol, squalene, fatty acids, or nucleolipids. These novel conjugates provide a means to enhance tissue targeting, cell internalization, endosomal escape, target binding specificity, resistance to nucleases, and more. We next describe those bioconjugated oligonucleotides approved for patient use or in clinical trials. Finally, we summarize the state of the field, describe current limitations, and discuss future prospects. Bioconjugation chemistry is at the centerpiece of this therapeutic oligonucleotide revolution, and significant opportunities exist for development of new modification chemistries, for mechanistic studies at the chemical-biology interface, and for translating such agents to the clinic.

## INTRODUCTION

The molecular biology central dogma describes the transfer of biological information stored in genes to proteins using biomacromolecules.[1] The biomacromolecule deoxyribonucleic acid (DNA) is found in all nucleated eukaryotic cells and it stores information via a specific sequence present in DNA. The first-step in this unidirectional process of genes to proteins is transcription whereby ribonucleic acid (RNA) is generated from DNA (i.e., transcription). RNA is then spliced into the nucleus to retain only the coding part of the RNA, namely the exons. The resulting messenger RNA (mRNA) is transferred to the cytoplasm, where the ribosome reads the information and produces a protein (i.e., translation). These biomacromolecules play key roles in all aspects of life from life cycle, pathogenesis, to healing and, thus, opportunities exist to control a biological outcome by intervening in transcription or translation.

In 1967, synthetic nucleoside derivatives were reported with the intent for specific base pairing with target sequences[2]. In 1978, Summerton developed methods for inactivating sequences through crosslinking[3], and Zamecnik and Stephenson[4,5] reported that short synthetic DNA fragments (called oligodeoxynucleotides, ODN) from the Rous sarcoma virus, possessing sequence binding complementary to RNA molecules, inhibited viral replication. This seminal discovery documented replication prevention of a viral RNA strain using a specific ODN - today known as antisense treatment[6]. Several studies have also elucidated many of the molecular mechanisms underlying different human and animal pathologies. Thus, altering the expression of pathological genes, whether genetic or modified by mutation, is of keen interest and technologies are being developed for such purposes.

Antisense DNA (ASO) and RNA interference (RNAi) are two very promising technologies for inhibiting specific genes. The former utilizes a DNA fragment complementary to the target mRNA sequence. The latter approach, first described by R. Jorgensen[7] and extensively studied by Andrew Z. Fire and Craig Mello,[8] uses small interfering RNAs (siRNAs) to target the mRNA. Antisense oligonucleotides induce different inhibition mechanisms, which can occur in the cytoplasm and in the nucleus (Figure 1). After transcription of the DNA into pre-messenger RNA, several post-transcriptional modifications are performed, including splicing, capping at the 5' end, or the polyadenylation of the 3' end. The mRNA is then translocated into the cytoplasm by exportins, where it is then translated into proteins by ribosomes. A synthetic oligonucleotide elicits its activity in two cellular compartments, the cytoplasm or the nucleus. After hybridization of the oligonucleotide on mRNA, two inhibitory mechanisms are possible: i) mRNA degradation or ii) translational repression. mRNA is degraded upon hybridization with an oligonucleotide following RNase H recognition. Antisense DNA can also inhibit the translation step *via* a steric hin-drance mechanism at the ribosome binding site[9] or by prohibiting the ribosomal motion along the mRNA.[10,11] Several mechanisms of inhibition are also possible in the nucleus, including the inhibition of post-transcriptional modifications, such as capping at the 5' end, or by binding to polyadenylation sites at the 3' UTR ends.[12,13] Additionally, antisense may also inhibit or promote the inclusion of exons to alter the splicing of pre-mRNA (Figure 1).

In principle, protein regulation is accomplished by the use of either agonist[14,15] or antagonist[16–18] strategies, with the caveat that the target RNA sequence is known. Because they are exogenous molecules, synthetic oligonucleotides are substrates for endo- and exo-nucleases, and thus exhibit short half-lives both *in vitro* and *in vivo*. For example, after intravenous (IV) injection in monkeys, phosphodiester oligonucleotide analogues are quickly degraded with a half-life of only 5 minutes.[19] In order to overcome this low stability, oligonucleotides are chemically modified. Sites of chemical modification include the base, ribose, and the phosphate linkage. The types of modifications vary from classical to non-classical bioisosteres.[20] Additionally, one chemical modifications does not always address the same half-life limitation(s), and often two or more modifications are combined to increase oligonucleotide stability. During the last few decades, many different modification chemistries have been developed (Figure 2). One of the first chemical modifications reported is the replacement of the phosphodiester by a phosphorothioate linkage (PTO) to minimize oligonucleotide degradation. Interestingly, this modification does not interfere with the recruitment of RNase H and target RNA cleavage. Structures of the different chemical modifications and the knockdown mechanisms are shown in Figure 2.[21–26] To improve the binding affinity and nuclease resistance, chemical modifications including the 2'-O-methyl (2'-OMe), 2′-O-methoxyethyl (2′-MOE), and/or 2'-fluoro (2'-F) modifications of RNA are inserted into the structure. To enhance the binding affinity, Locked nucleic acid (LNA) modifications are used where the conformational freedom of the ribose is restricted. More recently, Tricyclo-DNA (tcDNA) are reported as another constrained nucleotide featuring a three-ring scaffold. Importantly, these types of chemical modifications efficiently address the degradation issues. However, they do not improve the cellular uptake or targeting of the oligonucleotide to a specific site at the cellular, tissue, or organ level. Indeed, these hydrophilic poly-anions do not easily cross a physiologic barrier (e.g., skin) or a cell membrane.

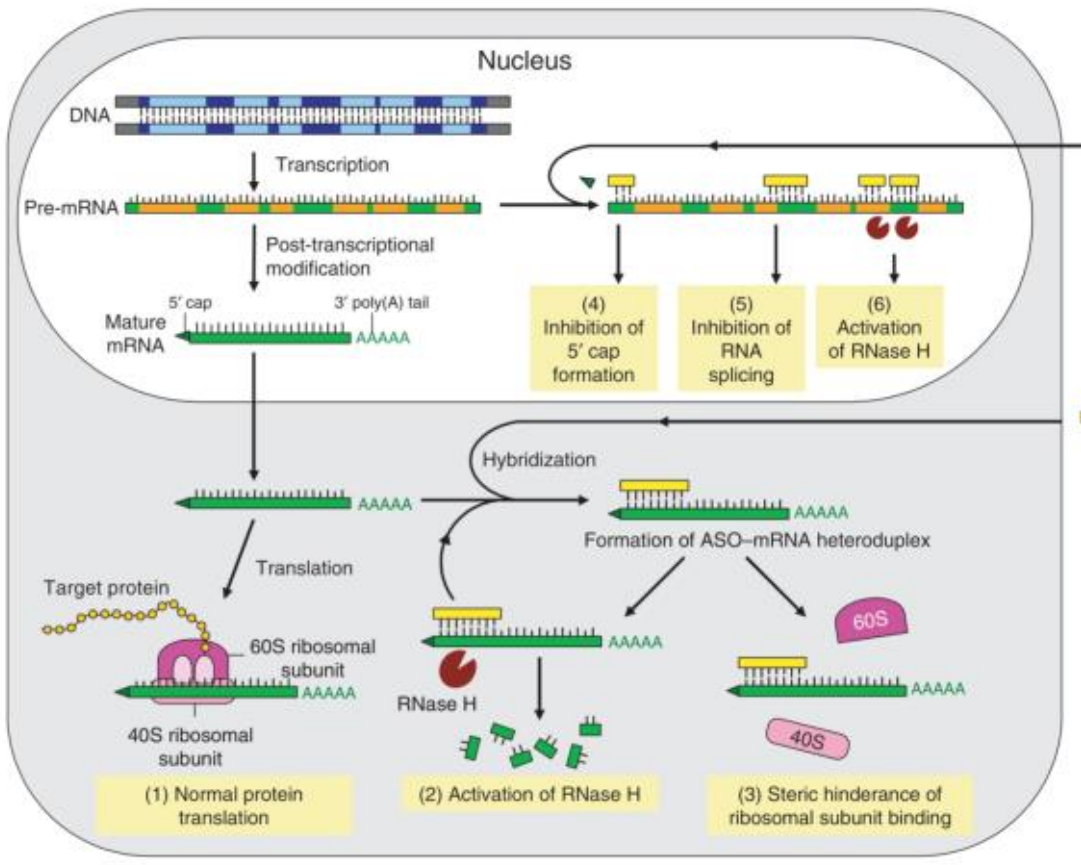

Figure 1: Mechanisms of antisense oligonucleotides (adapted from Chan J. H. P. *et al.* 2006 [12]) Normal expression of the gene and protein in the absence of ASO (1). After internalization, ASO hybridize to mRNA via complementary sequence pairing. The ASO/mRNA heteroduplex induces activation of RNase H and degradation of the mRNA (2) or ribosome blockage by steric hindrance (3). Both mechanisms result in inhibition of target protein production. Alternatively, penetration of ASO into the nucleus may lead to regulation of mRNA via inhibition of capping formation at the 5' end (4), inhibition of splicing (5) or the activation of RNase H (6).

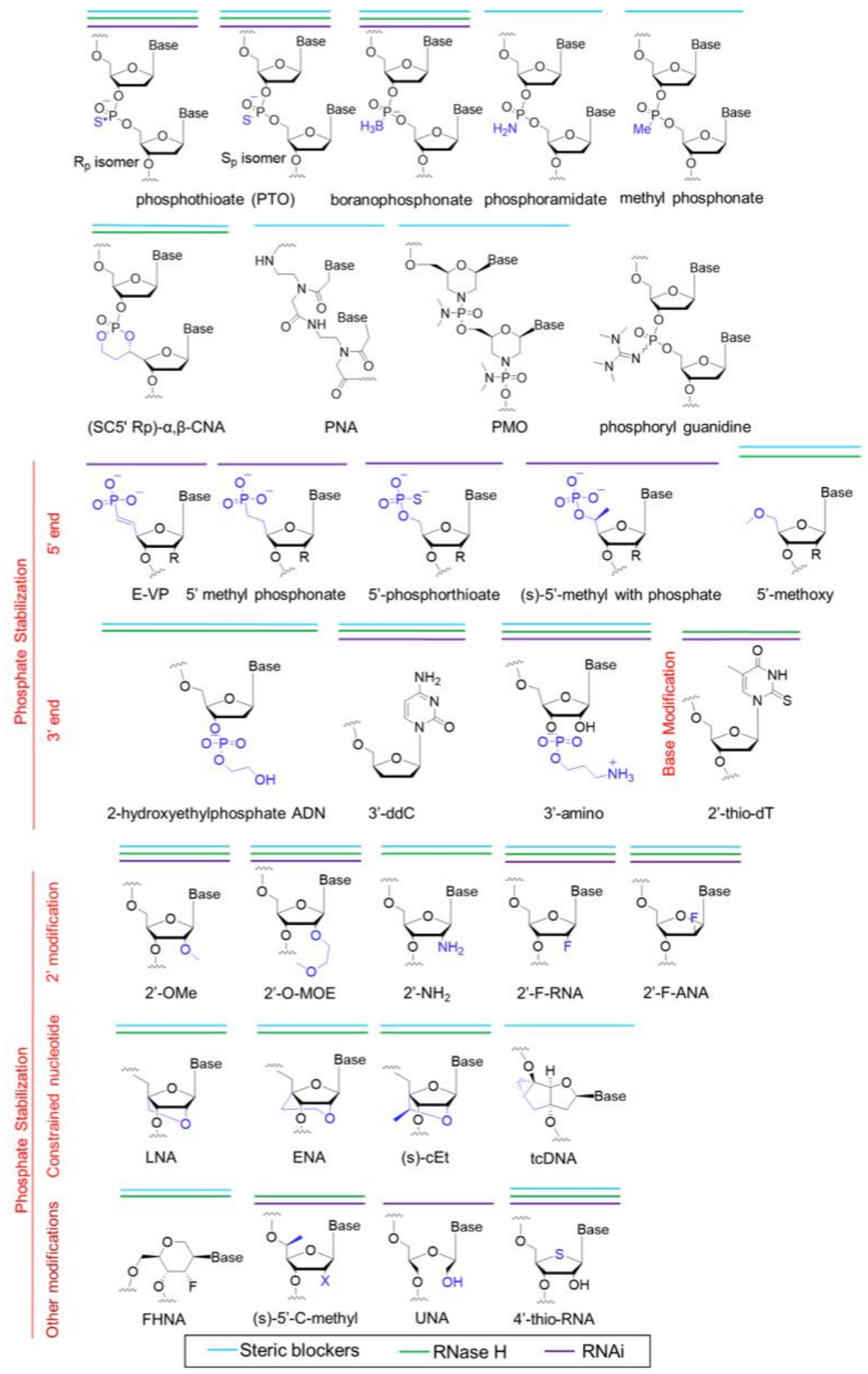

Figure 2: Chemical modifications of synthetic oligonucleotides (Adapted from Khvorova A. and Watts J. K. 2017 [21]) Chemical modifications are incorporated at the phosphate, ribose or at the 5' or 3' ends of the oligonucleotides. The type and the position of the modification induce three main mechanisms (steric blockers in green, activation of RNase H in blue and RNA interference (RNAi) in orange). In the case of 2' modification (2'-OMe, 2'-O-MOE, 2'$NH_2$, 2'-F-RNA and 2'-F-ANA), RNase H and RNAi mechanism are possible with a gapmer design. PTO: Phosphorothioate ; CNA : *Dioxaphosphorinane-Constrained Nucleic Acid* ; PNA : *Peptide Nucleic Acid* ; PMO : *Phosphorodiamidate Morpholino Oligonucleotide* ; E-VP : (E)-*VinylPhosphonate* ; ddC : dideoxyCytosine ; 2'-OMe : 2' O-Methyl ;2'-O-MOE : 2'-O-MethOxy Ethyl ; 2'-F RNA; 2'-F ANA : *ArabinoNucleic Acid* ; LNA : *Locked Nucleic Acid* ; ENA : *2'-O,4'-C-Ethylene-bridged Nucleic Acid* ; cEt : *S-constrained Ethyl* ; tcDNA : *tricycloDNA* ; FHNA : *Fluoro Hexitol Nucleic Acid* ; UNA : *Unlocked Nucleic Acid*.

Therapies must adhere to a wide range of stringent specifications in order to advance into clinical trials and succeed. An ideal oligonucleotide therapy is safe to administer, affordable to produce, exhibits a 2 year shelf-life, possesses an extended half-life in blood and serum (> 12 hrs), and inhibits intended targets effectively with low nonspecific or off-target interactions.[27] That said, differing disease indications will require fine tuning of the design requirements for optimal performance. As discussed below, a wide range of modifications to oligonucleotides improve one or many of these parameters, and the use of bioconjugation to oligonucleotides is primarily focused on improving the delivery of oligonucleotides to the cytosol or nucleus. Bioconjugates address a number of different oligonucleotide delivery challenges, such as improving biodistribution to a specific region or cell type (e.g., antibodies), promoting endosomal escape (e.g., CPP), increasing receptor-mediated transport (e.g., GalNAc), and/or increasing lipophilicity (e.g., cholesterol).

In this review, we discuss recent advances in the delivery of oligonucleotides. We briefly describe self-assembly approaches based on electrostatic, hydrophobic, or H-bonding interactions to give, for example, lipoplexes, and refer the reader to several comprehensive reviews on this topic.[28,29] We focus on modified bioconjugated oligonucleotides for delivery, including GalNAc, cell penetrating peptides, α-tocopherol, aptamers, antibodies, cholesterol, squalene, fatty acids, nucleolipids, and bis-conjugates. We also describe relevant linking chemistries between the oligonucleotide and the bioconjugate with respect to molecular design.. Additionally, the therapeutic applications of such bioconjugated oligonucleotides and outcomes from clinical trials are highlighted. Finally, we summarize the state of the field and discuss future prospects. Today, significant opportunities exist for development of new modification chemistries, for mechanistic studies at the chemical-biology interface, and for engineering systems for efficient delivery to the target site.

## *OLIGONUCLEOTIDE DELIVERY*

Cellular uptake is one of the key steps in order for oligonucleotides to elicit their biological activity, as the target mRNAs and the cellular machinery are located in the cytoplasm and in the nucleus. Several transfection agents have been developed and commercialized for the intracellular delivery of oligonucleotides such as oligofectamine 2000 (Invitrogen), JetPEI (Polyplus transfection) or K2 (Biontex). Many of these transfecting products are composed of cationic lipids, polyethylenimine, or DEAE-dextran, which self-organize and self-assemble in aqueous medium via electrostatic interactions with the negatively charged oligonucleotides. This strategy compacts the oligonucleotide into aggregates for subsequent endocytosis by the cell. Once inside the cell, the oligonucleotides are released into the cytoplasm. However, toxicity remains a disadvantage of these agents, as most of transfecting agents are cationic, but alternatives are being investigated.[30,31] Additionally, many of these complexes are not stable in the presence of serum, limiting or preventing their *in vivo* use.[32,33] Following intravenous administration, the resultant oligonucleotide/transfecting agent complexes accumulate primarily in the liver, due to their size.[34] Consequently, significant research efforts are dedicated to the delivery challenges in order to improve the transfection of nucleic acids.[35–37] For example, poly(2-dimethylaminoethyl) methacrylate, when mixed with oligonucleotides,[38] forms micelles of controlled sizes based on the N/P ratio. Upon adding an albumin coating to the surface, the cytotoxic effect of the polymer is minimized and cancerous cells are preferentially transfected relative to healthy cells.[39] Another strategy employs modified viruses, which internalize nucleic acids in cells.[40–42] However, this strategy can afford off-target toxicity since the oligonucleotides are not delivered in their synthetic form, but rather integrated into the viral genome. Lipid nanoparticles (LNP) are also being investigated as the non-viral transfecting cargo. However, the transfection efficiency is generally lower than that observed for viral transfection systems.[43]

Recently, LNPs loaded with siRNA targeting Polo-Like Kinase 1 (PLK1) protein, present in the triple negative breast cancer cell line (MDA-MB-231), have been modified with antibodies to target tumors. Biodistribution studies of labeled siRNA-LNPs demonstrated that antibody modified LNP (antibody against heparin-binding EGF-like growth factor, αHB-EGF) effectively delivered siRNA to tumor tissue in mice. Interestingly, the PLK1 protein expression was inhibited after intravenous injection of the LNPs and tumor growth was decreased. These results indicate that antibody modified LNPs loaded with siRNA are a promising therapeutic approach for breast cancer.[44] Another recent article investigated the bioconjugation of antibodies to LNP for targeting endothelial cells lining the vascular lumen.[45,46] These new particles were non toxic both *in vitro* and *in vivo* as well as preferentially accumulated in the lungs. In contrast, the same nanoparticles without conjugated antibodies quickly accumulated into the liver, indicating that targeting moieties (antibodies/LNP bioconjugate) can avoid the hepatic uptake with endogenous serum protein like Apo-E.[47]

Alternatively, bioinspired molecules such as nucleolipids (NL) are being used to construct LNPs. NLs self-assemble to form unique supramolecular structures,[48–52] and the LNPs based NLs loaded with nucleic acids successfully transfect plasmid DNA, siRNA, and antisense oligonucleotides to a number of different cell lines: human breast adenocarcinoma MCF-7 cells, human liver (HepG2), mouse fibroblast (NIH 3T3), Chinese hamster ovarian (CHO) cells, and human prostate cancer (PC-3) cells.[51,53–55] Furthermore, these LNPs can be further modified to be stimuli-responsive, such as responding to changes in pH to enhance the delivery of nucleic acids.[56] The above examples are representative and by no means comprehensive, as there are many formulations described for nucleic acid vectorization (Figure 3), and the reader is referred to several comprehensive reviews on the subject.[57,58]

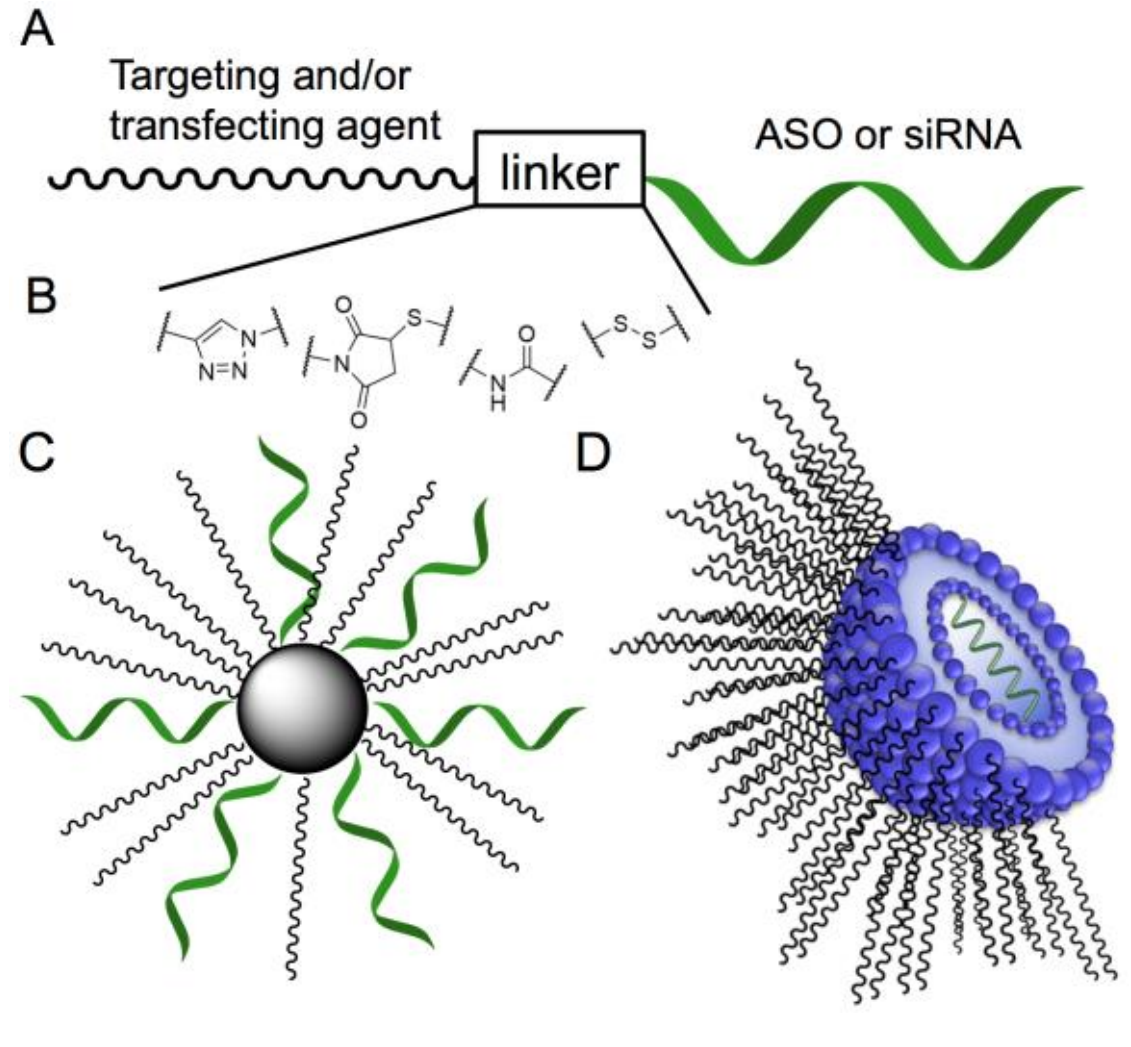

Figure 3: Strategies for the delivery of ASO and/or siRNA. Adapted from Järver *et al. 2012* [57] A: Formation of a complex between a biomolecule or a polymer and the oligonucleotide *via* electrostatic interactions and / or the hydrophobic effect, B: Formation of liposomal supramolecular structure with targeting or transfecting agent (PEG, peptide, CPP, protein, lipid glycoconjugate, etc.), C: Covalent conjugation with the oligonucleotide by a stable or cleavable bond.

The re-targeting of nucleic acids using viral vectors was investigated by Reynolds *et al.* in the 2000's.[59,60] Viral vectors are attractive candidates for *in vivo* gene delivery because the infection efficacy is higher compared to other non-vial approaches. For example an adenovirus vector was prepared containing both a Fab fragment of an anti-Ad5 knob antibody and the anti-ACE monoclonal antibody mAb 9B9. This bi-specific conjugate exhibited enhanced pulmonary distribution by a synergic effect (transductional and transcriptional).[59,60] The major drawback of using this vector is sequestration by Kupffer cells into liver tissue.

## BIOCONJUGATED OLIGONUCLEOTIDE DELIVERY

The conjugation of specific molecules to oligonucleotides is a promising therapeutic approach for nucleic acid based drugs. Consequently, bioconjugates are of increasing presence in the pharmaceutical development pipeline. The major advantages of working with a bioconjugate include: 1) a new chemical entity; 2) of defined composition; and, 3) synthesized using chemical methods as opposed to bioprocesses. These attributes, unlike formulations with polymers or other transfecting reagents[61] that give heterogeneous mixtures requiring extensive multi-pronged characterization analyses, will facilitate translation to the clinic. Additionally, these covalently conjugated molecules play one or more roles in recognition, targeting of cells or tissues, cellular internalization, and pharmacokinetics.

The review of covalently conjugated oligonucleotides necessitates discussion on appropriate and effective linkers and linking chemistries, as interference strategies can be stymied by functionalization(s)[62]. Both cleavable and stable linkers are successfully used. Bio-orthogonal "click" chemistry approaches such as akyne-azide, and thio-maleimide are two common approaches to form stable linkages with highly specific reactions. Cleavable bonds such as reducible disulfide linkages, esters (cleavable through hydrolysis or esterases), phosphodiesters (cleavable through nucleases) and peptides (cleavable through proteases) are also being utilized and examples are discussed below. Additionally, alternatives such as end functionalization (5' or 3'), and other cleavable strategies will be discussed in the context of different conjugated moieties.

***PEG modification.*** Poly(ethylene glycol) (PEG) was first conjugated to oligonucleotides more than two decades ago by different academic groups, including Bonora,[63] and Burcovich[64] in order to improve stability, avoid rapid degradation, and enhance cellular uptake.[65] In 2003, Ji Hoon Jeong *et al.* described PEG attached to an antisense oligonucleotide sequence (*c-myb*) to afford a surrounding corona, which would prohibit interactions with plasma proteins and increase aqueous solubility.[66] Intracellular uptake is achieved due to the formation of a micellar system resulting from the combination of PEG conjugates with fusogenic cationic peptides. These formulations exhibit higher antiproliferative activity against smooth muscle cells compared to the unconjugated *c-myb* antisense.

***GalNAc modification.*** The GalNAc (N-acetylgalactosamine) modification, introduced by Nair *et al.*, lead to hydrophilic glycoconjugates[67] (Figure 4), increasing the cellular internalization in the liver due to binding to the ASialoGlycoProtein Receptor (ASGPR). These lectin type C receptors are highly expressed on the plasma membrane of hepatocytes and both galactose and GalNAc are the preferred ligands for these receptors. After binding to ASGPR, the complex is quickly internalized to form early endosomes *via* a clathrin dependent mechanism.[68] As a result, ASGPR targeting conjugates are readily delivered to the liver through intraveneous injection.

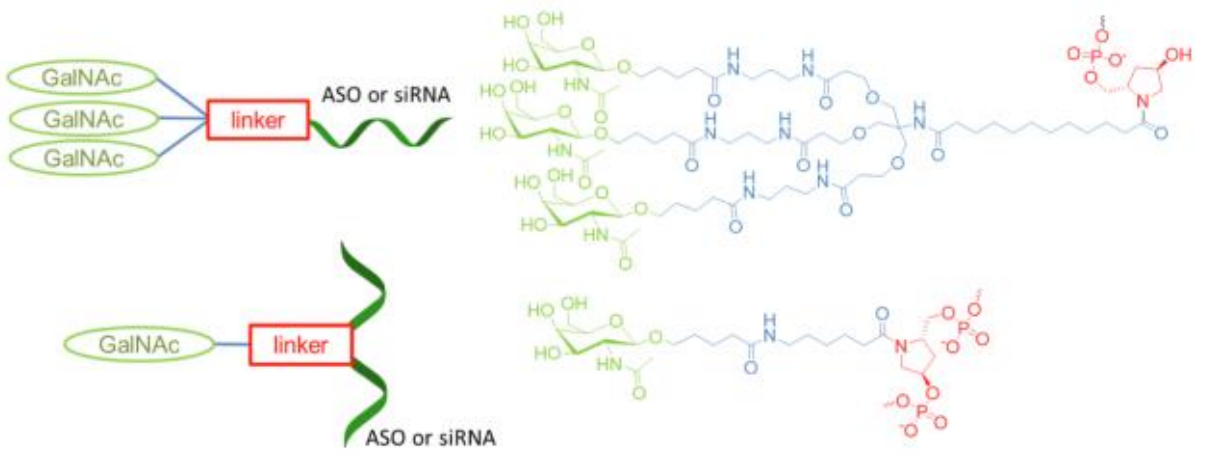

Figure 4: Molecular structures of GalNAc based conjugates, adapted from Matsuda *et al. 2015* [72] Diagram showing an oligonucleotide coupled to GalNAc (left) and molecular structure (right). The triantennular GalNAc is represented in the upper part, the GalNAc incorporated in sequence in the lower part.

Once inside the cell, the oligonucleotides must escape from the endosome intact to elicit their bioactivity. At neutral pH, the dissociation constant between these modified oligonucleotides and their receptor is very low (Kd in the nanomolar range), indicating a strong association with ASGPR receptors. Once inside the endosome, the acidic pH allows for dissociation of the ligand-receptor complex, resulting in the return of ASGPR to the membrane (Figure 5) and subsequent release of the ligand. The GalNAc modification facilitates cell internalization. Finally, the GalNAc modified oligonucleotides escape the endosome.[69]

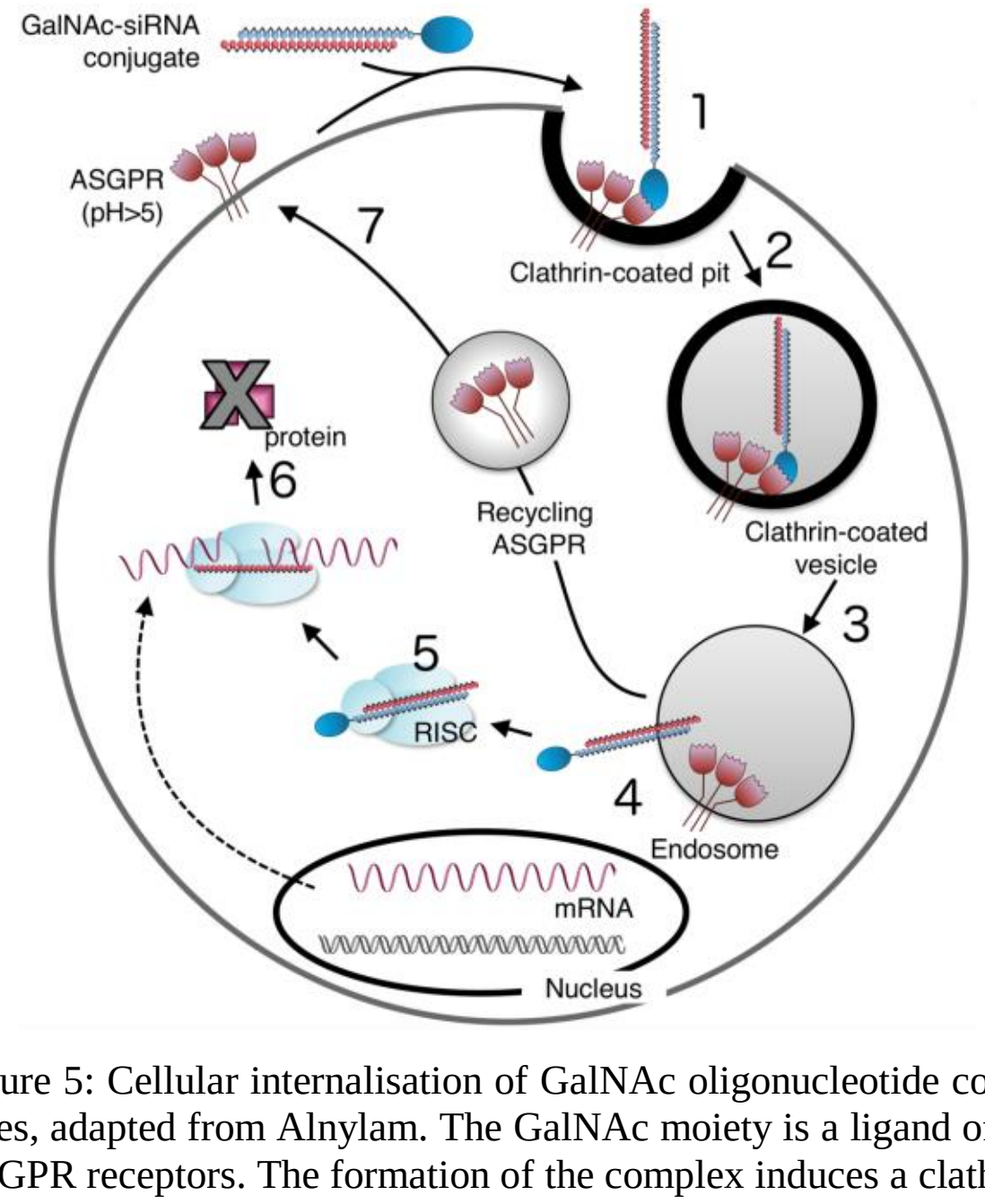

Figure 5: Cellular internalisation of GalNAc oligonucleotide conjugates, adapted from Alnylam. The GalNAc moiety is a ligand of the ASGPR receptors. The formation of the complex induces a clathrin-dependent endocytosis mechanism. Acidification dissociates the siRNA-releasing substrate / receptor complex (3) into the cytoplasm (4). RNAi-mediated degradation of the targeted mRNA mechanism (5). Inhibition of protein translation (6). ASGPR receptors are recycled in the plasma membrane (7).

To prepare these bioconjugates, the GalNAc modification is introduced into the oligonucleotide either during DNA synthesis as a non-nucleoside monomer or at the end of the oligonucleotide synthesis as a "triantennular" form (Figure 4). The synthesis of such oligonucleotide conjugates is compatible with standard automated solid phase synthesis, where the GalNac modification is attached either at the 3', 5' extremities or within the oligonucleotide sequence.[70] The non-nucleoside monomer is attached through a phosphodiester linkage during oligonucleotide synthesis as a synthetic phophoramidite. All of the different chemical modifications (2'-OMe, 2'-F, PTO), which can be inserted within the same sequence, are compatible with the GalNac conjugation approach. Manoharan *et al* reported that these modifications improve the *in vivo* stability of siRNA−GalNAc conjugates.[67] Interestingly, after subcutaneous injection, the siRNA−GalNAc conjugates distribute to the liver with subsequent gene expression decrease of the targeted mRNA, thus enabling this approach for treating a wide range of liver diseases.[71,72] The pharmacokinetics of GalNAc-siRNA conjugates were evaluated in monkeys[73] and it was found that phosphorothioate linkages in the 5'-region of siRNA−GalNAc strands minimally affect plasma pharmacokinetics but result in a greater liver exposure. This provides a prolonged duration of gene silencing due to improved metabolic stability. The technology was also developed for ASOs targeting apolipoprotein with a successful improvement (10-fold) of inhibition effect *in vivo*.[74,75] In this study, the active ASO is liberated only after internalization into cells and not in the plasma, demonstrating that ASO-GalNac conjugates are a functional prodrug. Many GalNAc conjugated oligonucleotides are currently in preclinical and clinical trials.[76] For example, Fitusiran[77,78] is a siRNA developed by Alnylam for the treatment of hemophilias. The collaboration between Ionis and Akcea yielded ASO IONIS-APO (a) LRX for the treatment of hyperlipoproteinemia and cardiovascular diseases. Additionally, two anti-miRNAs are being developed by Regulus to treat viral diseases, such as hepatitis C (RG-101), and a third is being developed in collaboration with AstraZeneca for the treatment of type 2 diabetes (RG-125).[76]

***Peptide Sequences: Cell Penetrating Peptide* (CPP) and RGD.** CPPs are short peptidic sequences, generally not exceeding thirty residues, which possess the ability to cross a cellular membrane and facilitate endosomal escape (Figure 6). CPP are classified into three categories: 1) protein-derived peptides; 2) chimeric peptides formed by the fusion of two natural sequences; and, 3) synthetic peptides based on structure/function studies.[79] These short peptides are ligands for specific receptors that facilitate cell internalization by endocytosis and destabilization of endosomes compartments. For example, Lönn *et al.* speculated that the PTD/CPP-EED domains enhance cellular delivery by insertion of a hydrophobic patch into the lipid bilayer at a critical distance (18 bonds) from the delivery domain. This resulting concentration in the endosome results in a strong localized membrane destabilization leading to enhanced escape into the cytoplasm.[80] All CPPs do not exhibit the same penetration mechanism, membrane leakage rates, or toxicity. In 2007, a CPP ((RXR)4 (X = 6-aminohexanoic acid)) conjugated to a PMO oligomer was investigated.[81] The CPP-PMO conjugates did not show toxicity below a dose of 15 mg/kg after intravenous injection (bolus). However at a dose of 30 mg/kg, animal body weight decreased while at a dose of 150 mg/kg, the animals were lethargic with elevated levels of creatinine. Blood biochemisty analysis of albumin, electrolytes, and bilirubin showed constant levels at all doses.[81]

A major design issue for these conjugates is the complexation between the anionic oligonucleotides and the cationic

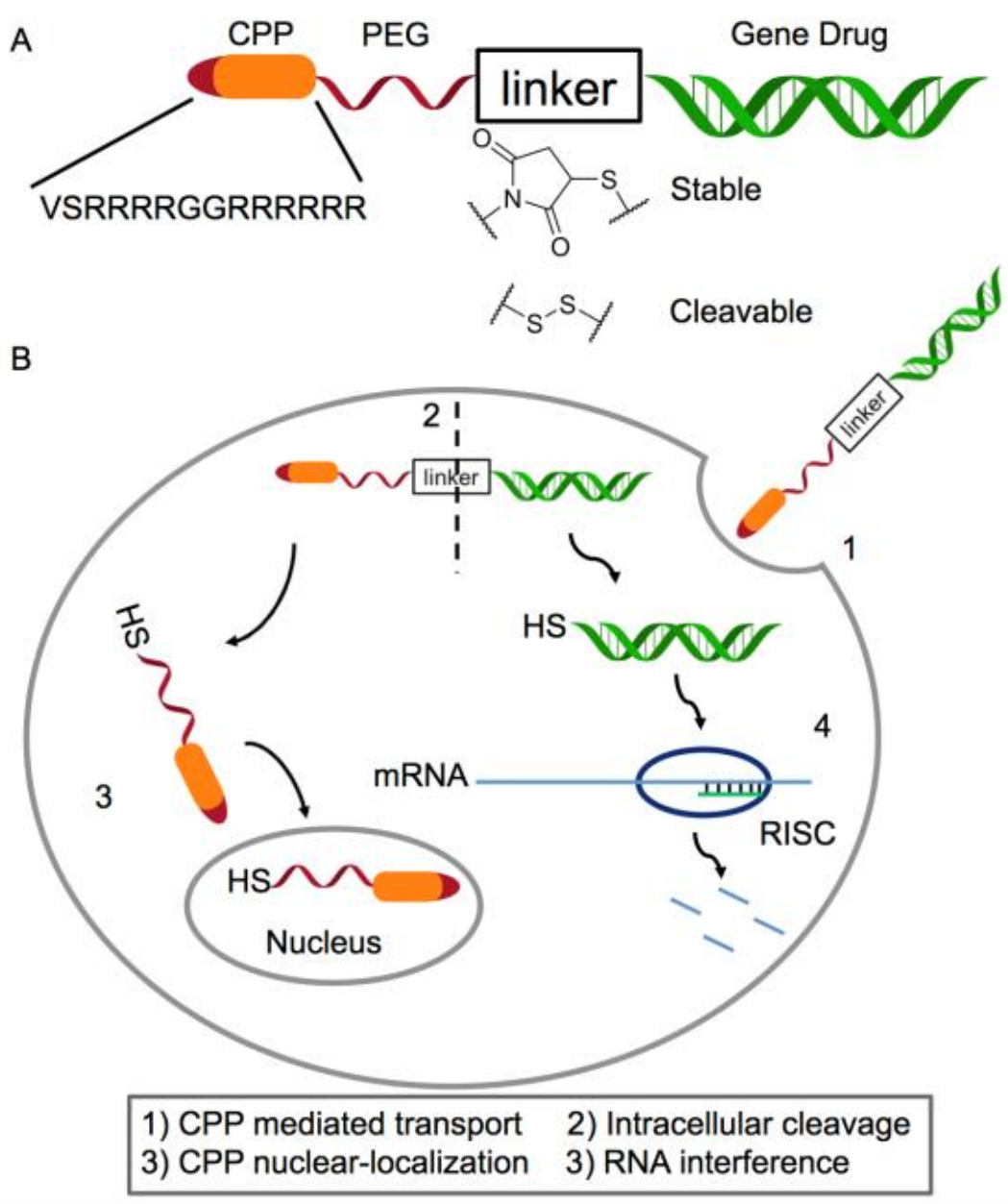

Figure 6: CPP grafting strategy on siRNA. A: Noncleavable oligonucleotide CPP conjugate. B: Representation of the cleavable conjugate by reduction of the disulfide bridge: (1) Bioconjugate enters into the cell thanks to CPP. (2) Cleavage of the disulfide bridge allows detachment of the siRNA from the CPP moiety. (3) CPP enters into the nucleus. (4) siRNA elicits its inhibitory activity by an RNAi mechanism. Adapted with permission from Ye, J., Liu, E., Gong, J., Wang, J., Huang, Y., He, H., and Yang, V. C. (2017) High-Yield Synthesis of Monomeric LMWP(CPP)-SiRNA Covalent Conjugate for Effective Cytosolic Delivery of SiRNA. *Theranostics* ***7***, 2495−2508.[83] Copyright 2017, Ivyspring.

CPPs. In a previous review, Steven Dowdy focused on the potential use of CPP to deliver siRNA, and the reader is referred to this manuscript.[82] The endocytosis-mediated uptake of such peptides depends on three important steps: cell association, internalization, and finally endosomal escape. Dowdy *et al*. also discuss the array of different cargos that have been delivered by cationic PTDs/CPPs as well as cellular processes and biological responses that have been modulated. CPPs share many common features responsible for efficient activity including positive amino acids (arginine and lysine) and hydrophobic residues (tryptophan and phenylalanine).[83,84] Alternatively, non-ionic oligonucleotides (PMO, PNA, phosphoryl guanidine) are good candidates to bypass this charge-charge complexation requirement. Such systems are used successfully for exon skipping.[85–87] Conventional formulations that carry oligonucleotides are composed of CPP via physical mixing, but the conjugation of these molecules to the oligonucleotide by a covalent bond provides additional benefits. The chemical insertion of multiple peptides on the same oligonucleotide improves the sensitivity of the conjugate for its receptor. In the case of siRNA conjugated to cyclic RGD, the internalization mechanism is achieved by caveola-dependent endocytosis.[88,89] In another study, the bombesin peptide, coupled to the 5' end of a splice-shifting oligonucleotide (SSO), corrects splicing of an aberrant intron. This fourteen amino acid peptide, which is a neurotransmitter acting on the G protein receptors, facilitates oligonucleotide delivery to the nucleus of transfected prostate cancer PC3 cells.[90–92] Ye *et al.*[83] demonstrated that a 3' CPP covalently conjugated to siRNA increased their intra cellular delivery (Figure 6). The report investigated CPPs with different conjugation chemistries with a PEG space, one cleavable disulfide linkage, and another through a thiol-maleimide coupling. The incorporation of a stable versus cleavable linkage led to differences in knockdown and efficacy.

The linear tripeptide RGD and its cyclic-RGD analogs are well known peptides for cell targeting. These peptide ligands bind αvβ3 integrins which are overexpressed in several cancers, including melanomas,[93] ovarian cancer,[94] and breast cancer as well as in metastasis.[95] Grafting such ligands to therapeutic oligonucleotides provides a means for targeting and efficient delivery. However, a drawback of using the RGD as a targeting moiety is that the αvβ3 integrin is expressed on many cell types as well as platelets, and, as such off-target effects and toxicities are of concern. Although this class of peptide-modified bioconjugates is being actively investigated for the delivery of nucleic acid *in vitro*, positive results still remain to be confirmed *in vivo*. Additionally, biodistribution and resistance to proteolytic degradation are not reported.

***α-Tocopherol.*** α-Tocopherol, commonly known as vitamin E, is used for the treatment of hypercholesterolemia. This liposoluble moiety is often coupled to oligonucleotides to facilitate cell membrane interactions.[96] For example, Yutaro Asami *et al.* coupled tocopherol to a siRNA at the 5' end of the 29-nucleotide antisense strand for reducing apolipoprotein B (ApoB) levels. Improved inhibitory activity is observed relative to the parent cholesterol derivative. In fact, only 2 mg/kg is needed to significantly reduce ApoB mRNA in mouse liver after intravenous injection. In similar conditions, the amount of cholesterol conjugate required to observe the same activity was 50 to 100 mg/kg.[97] Additionally, α-tocopherol, coupled to an ASO, increases endogenous gene inhibition in a mouse liver compared to an unconjugated ASO.[98] Simply conjugating tocopherol to the oligonucleotide structure did not induce an inhibitory effect, and a spacer of several (4-7) Unlocked Nucleic Acids (UNA) was required between the oligonucleotide 5'extremity and tocopherol to maintain biological activity. In this study, the oligonucleotides within the ASO selected were Gapmers, oligonucleotides containing a mixture of chemically modified nucleotides (LNA, 2'-O-MOE, for example) at each extremities and a central sequence of DNA (the "gap"), allowing the RNase H cleavage after forming the mRNA/DNA complementary heteroduplex. The UNA residues, on the spacer, are cleaved in the cell affording the ASO using the RNase H mechanism for inhibition of mRNA. The inhibitory efficacy increased by 3.5 fold compared to the intravenous injection of non-conjugated oligonucleotides. Importantly, tocopherol improved the pharmacokinetics of ASO affording accumulation in the liver (9-14 □g/g and 2-5 □g/g were found in the liver 6 h after injection at a dose of 3 mg/Kg for the conjugated and unconjugated ASO, respectively), unlike the unconjugated ASO, which distributed primarily to the kidneys (1-2 µg/g and 12-16 µg/g were found in the kidneys 6 h after injection at a dose of 3 mg/Kg for conjugated and unconjugated ASO, respectively).

***Aptamers.*** Aptamers are oligonucleotide sequences (DNA or RNA) possessing a three-dimensional structure for biological recognition. Using a method called SELEX (Systematic Evolution of Ligands by EXponential enrichment) specific oligonucleotides are identified that bind a target with high affinity ($K_d$ in the range of nanomolar to picomolar). Sullenger *et al.*[99] used this approach to select an aptamer capable of targeting the Prostate-Specific Membrane Antigen (PSMA) receptor at the surface of prostate cells. The designed RNA molecule contained both the aptamer and siRNA moieties. This chimeric aptamer-siRNA features a dsRNA (siRNA) motif that is a substrate for the intracellular Dicer enzyme, an endoribonuclease. As a proof of concept, LNCaP and PC-3 prostate cells were treated with the chimeric aptamer-siRNA containing a therapeutic siRNA sequence that targets genes overexpressed in human tumors (PLK1 and BCL2). Successful knockdown resulted in a decrease in cell proliferation and an increase in apoptosis. As a control, an aptamer chimera bearing two point mutations exhibited an absence of specificity for its target, indicating the specific binding between the aptamer and its receptor. In another study, a siRNA targeting the tat/rev viral genes was conjugated *via* a short polynucleotide link to an aptamer that recognizes the glycoprotein gp120 of the HIV viral envelope. This conjugate suppressed HIV infection in human cell lines and in humanized mouse models.[100–102] Aptamer-siRNA chimeras are of significant interest and the reader is referred to a recent review summarizing the state-of-the-art.[103] The additional advantages of such chimeras include lower production costs, low batch-to-batch variation, longer shelf-life and little-to-no immunogenicity and toxicity compared to antibody conjugates (discussed below).

Given that the three-dimensional structure is required for recognition and that aptamers are degraded by nucleases, significant research efforts are directed at modifying nucleotides to increase the stability without losing target recognition. There are several locations and types of modifications being investigated including: the terminals of nucleic acids, the phosphodiester linkage, the sugar ring and modifications on the bases, as well as 3' end capping with inverted thymidine and PEG conjugation.[104]

***Antibodies.*** Monoclonal antibodies are known for their recognition properties of biological targets. In 2005 Lieberman *et al.*[105] reported the use of a monoclonal antibody conjugate for the delivery of siRNA in order to inhibit the protein production of HIV. The antibody was conjugated to the ASO oligonucleotide *via* a click reaction involving an azide-modified antibody and a cyclooctyne ASO. This strategy was further developed by Satake *et al.*[106] to conjugate the anti-CD22 antibody to the MXD3 oligonucleotide. This oligonucleotide targets a dimeric protein (MXD3) associated with MYC, a transcription factor responsible for the survival of B cell precursor cells in acute lymphoblastic leukemia. The conjugate induces inactivation of the MXD3 protein responsible for apoptosis *in vitro* in leukemic cells as well as affords cytotoxicity in normal B cells but not in hematopoietic cells, including stem cells. The *in vivo* results are promising as only a dose of 0.2 mg/kg twice a week for 3 weeks is required to double survival (median 42.5 vs. 20.5 days). Based on these encouraging results, antibody-ASO conjugates are a new therapeutic option to decrease the doses and to favor the accumulation of ASOs at designated locations. However, the selection of antibody or antibody fragment for targeting a specific cell type is challenging. For example, Genentech published an article describing the investigation of antibodies conjugated to ASO by a maleimide linkage, for targeting prostate carcinoma cells after systemic administration.[101] This challenge is further complicated by the fact that the route of antigen internalization also affects silencing. Although antibody conjugates reach the intended target site, the subsequent steps of cellular internalization and endosomal escape are also required for efficacy.[107]

***Cholesterol.*** Lipid moieties, and particularly cholesterol derivatives, are often conjugated to oligonucleotides for delivery purposes (Figure 7). Oligonucleotides modified with cholesterol are recognized by high and low-density lipoproteins (HDL and LDL) *in vivo* and internalized *via* cholesterol binding receptors. After intravenous or intraperitoneal injection, these oligonucleotide conjugates escape renal clearance, thus greatly impacting their pharmacokinetics by extending the time that the nucleic acids remain in the plasma. A number of studies have demonstrated the advantage of coupling a cholesterol moiety to the oligonucleotide in order to reduce its renal clearance. Also, cholesterol conjugation increases cellular uptake of oligonucleotides. Godeau *et al.* reported improved *in vitro* cellular internalization of an oligonucleotide functionalized with cholesterol *via* a "click chemistry" approach. Additionally, these conjugates inhibited viral translation of hepatitis via an IRES blocking mechanism.[108] Additional studies in the KB-8-5 cell line showed that the cholesterol moiety allows endosomal escape without the use of any additional transfecting reagents. After 4 hours, the siRNA conjugates accumulate in the intercellular space. After 24 hours, the distribution became homogeneous with an accumulation of the oligonucleotides in the cytoplasm of the cells.[109,110]

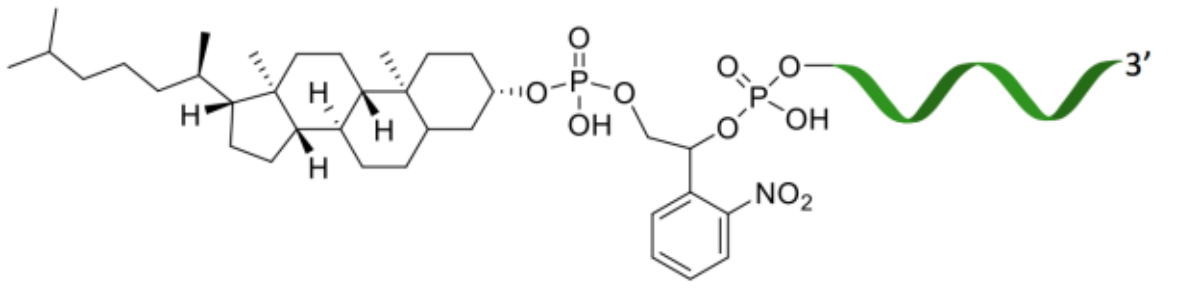

Figure 7: Molecular structure of a photocleavable cholesterol-conjugated oligonucleotide, adapted from Yang J. *et al.* 2018[111]

From a design perspective, the spacer between the oligonucleotide and cholesterol is crucial to maintain the biological activity. Directly grafting the cholesterol to the 5' oligonucleotide end reduced the inhibitory activity of the nucleic acids (siRNA or ASO).[109,111] In another study, ApoB was targeted to the liver after a daily intravenous injection of siRNA-cholesterol conjugates at a dose of 50 mg/kg.[112] Additionally, the Huntingtin protein was inhibited after a single intrastriatal injection of a cholesterol-coupled siRNA. These two examples demonstrate that siRNA-cholesterol conjugation improves the delivery into different cell types other than liver cells.[113] A photo-responsive cholesterol-siRNA conjugate (on both sense or antisense strands) is recently reported containing a photo-cleavable linkage (Figure 7)[114]. The orthonitrobenzyl group photo-cleaves in response to 365 nm UV light to yield the native 5' end. Three different cholesterol derivatives were prepared to inhibit 3 different genes (2 exogenous, luciferase and GFP; and an endogenous, Eg5) in hepatocarcinoma HepG2 cells. In the absence of light, this conjugation allows the internalization of the amphiphilic cholesterol-siRNA without inhibiting the target. In the presence of a light stimulus, the link between the cholesterol and the siRNA is cleaved, and the inhibitory activity of siRNA is introduced.[111]

Cholesterol is conjugated to oligonucleotides at different locations including the 5' or 3' end, as well as within the sequence. A recent study reported six constructs conjugated with cholesterol covalently bound either at the ends or after thymidines within the sequence. The oligonucleotide selected was a Gapmer with 2 and 3 LNAs at both the 5' and 3’ ends, respectively. Higher *in vitro* activities are obtained when the cholesterol is inserted either at the 5' end or within the sequence. *In vivo*, the cholesterol conjugates accumulate in the liver independent of the lipid conjugation location (3’, 5’-ends or within the sequence). Furthermore, a regular phosphodiester linkage exhibited superior *in vivo* performance in terms of biological activity (by a factor of 5) compared to the analog phosphorothioate as a link between the lipid and the ASO. Nakajima *et al.* suggests that the unmodified phosphate is cleaved in the liver freeing the non-conjugated ASO.[115] In 2007, Moschos *et al.* covalently attached the cholesterol moiety to the oligonucleotide sequence *via* a disulfide linker that is cleaved under intracellular reductive conditions.[116] In this study, the authors suggest that conjugation to cholesterol extends but does not increase siRNA-mediated mRNA knockdown in the lung (MAP kinase mRNA in mouse lung).

***Squalene.*** Squalene is a triterpene molecule and biosynthetic precursor of cholesterol found in plants and animals. Squalene is often conjugated to the 3'-end of the sense strand of siRNA via a thiol-maleimide coupling. Due to the resulting amphiphilic character of the squalene-oligonucloetide conjugate, these conjugates spontaneously formed spherical assemblies of 165 ± 10 nm diameter with a zeta potential of − 26 mV in aqueous media. These nanoparticles show enhanced oligonucleotide stability in the presence of serum and higher catalytic potency *in vitro* and *in vivo* against the targeted RET/PTC1 mRNA compared to the unmodified siRNA. Up to 70% inhibition of the tumor growth is observed after 19 days following intravenous injections of 2.5 mg/kg modified siRNA at day 0, 2, 4, 7 and 10.[117,118]

***Fatty acids.*** Currently in clinical trial (phase 2) for the treatment of myelofibrosis, GRN163L is a telomerase targeting, antisense oligonucleotide of thio-phosphoramidate with a covalently linked palmitic acid at the 5'-end (Figure 8).[119–121] The palmitic acid moity is conjugated though a PTO linkage to the backbone of the ASO. Telomerase activity is up-regulated in numerous cancers and, thus, is a viable target for treatment. Telomere shortening and lower cell viability are observed after inhibition of telomerase activity in cancer cell lines. The $IC_{50}$ values ranged from 50 to 200 nM for telomerase expression in 10 different pancreatic cell lines in the presence of GRN163. Telomerase reactivation and elongation are also observed when the antisense is no longer present, indicating that sequences targeting the RNA template region of telomerase are effective inhibitors of telomerase. GRN163L is a potential adjuvant for cancer treatment as it halts the progression of the tumor by restoring a normal, correct phenotype for the cancerous cells.

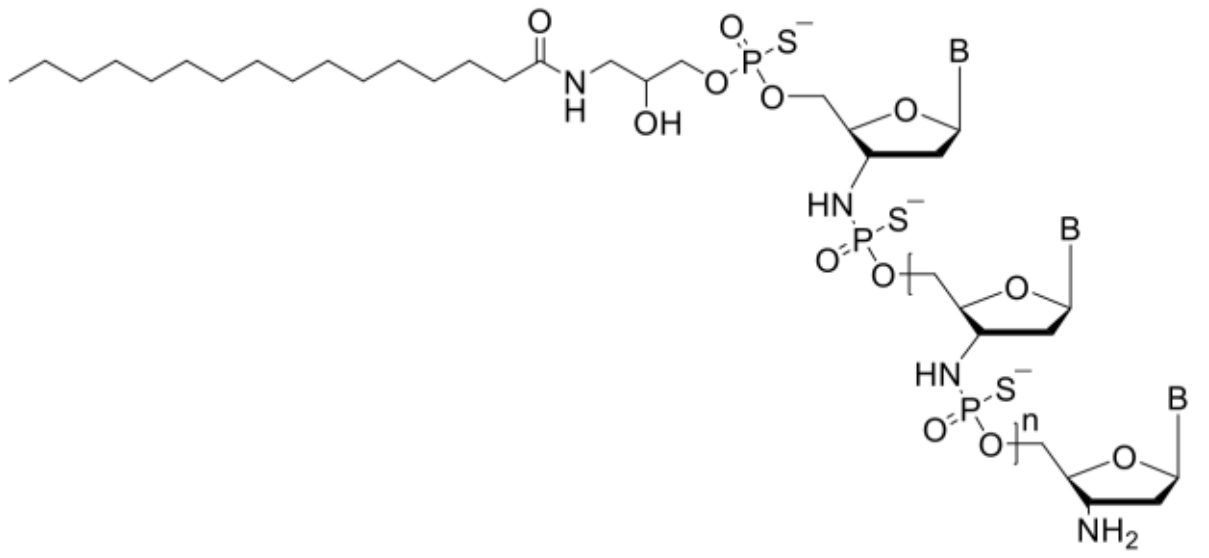

Figure 8: Molecular structure of GRN163L. The lipid oligonucleotide GRN163L, 5’ – Palm – TAGGGTTAGACAA – $NH_2$ – 3’, with a thio-phosphoramidate inter-nucleosidic linkage (top) and its structural formula (down) from the Chemspider website.

The *in vitro* inhibition of telomerase activity by ASO is improved with the palmitoyle modification. An inhibition factor ranging from 1.4 to 39 is observed depending on the cell type (cervix, glioblastoma, hepatocarcinoma, lungs, melanoma, myeloma, ovary and prostate). *In vivo* efficacy (up to 56% inhibition of telomerase activity after 24h compared to the uncoupled antisense or PBS) is also observed in a murine xenograft model following intravenous injection (50 mg/kg) in the absence of any additional transfecting reagent.[119] Additionally, investigations reveal a synergistic effect when GRN163L is co-administered with trastuzumab, a recombinant monoclonal antibody targeting HER2+ receptors in breast cancers.[121] The composition of the lipid (oleic vs palmitic acid in the lipid oligonucleotide conjugate) was also evaluated, but with no noticeable difference in efficacy.

***Ketal nucleolipid conjugate.*** Nucleolipids, first reported by our groups, are also being investigated.[48,52] For example, the ketal uridine phosphoramidite with two alkyl chains at the 2' and the 3' positions, anchors oligonucleotides to liposomal membranes[122]. These lipid-oligonucleotides are amphiphilic with micellar aggregation properties that are used as a functional reservoir for hydrophobic drugs. The drug payload of the micelles is released in the presence of the oligonucleotides complementary to the lipid-DNA conjugate.[123] Similarly modified ASOs are investigated for the treatment of prostate cancer.[124] An antisense strategy based on targeting the Translational Controlled Tumor Protein (TCTP) is described for the treatment of Castration resistant prostate cancer (CRPC), a disease currently with a very low survival rate (median survival reported between 9 and 30 months).[125–127] TCTP is minimally expressed in healthy prostatic tissue, moderately in prostatic cancer cells, and overexpressed in CRPC cells. Therefore, designing an inhibitory sequence is key for this promising strategy for the treatment of CRPC. An ASO sequence of 20 nucleotides with PTO backbone was selected to impede TCTP expression. Normally, this ASO sequence requires the assistance of a vectorization agent such as oligofectamine to be efficient *in vitro*. The conjugation of the ASO with the ketal nucleolipid enables the inhibition of TCTP in the absence of transfecting agent *in vitro* and *in vivo.*[124]

***Bis-conjugates.*** Multi-functionalized, bis-conjugated oligonucleotides are also being investigated. Tajik-Ahmadabad *et al.*[128] designed an amphiphilic CPP-modified ASO (in the PMO series) capable of self-assembly. The ASO is functionalized by coupling a 3-maleimidopropanoic acid moiety to the free secondary amine group at the 3' end. This maleimido-PMO was purified by HPLC method. The fatty-acid modified peptide was prepared by coupling myristic anhydride at the N-terminus of the resin-bound ApoE. After deprotection and resin cleavage, the ApoE peptide with C-terminal cysteine was conjugated to the ASO by thiol-maleimide click reaction. The spontaneous self-assembly of this ASO occurs when myristic acid is covalently attached to the N-terminus of the peptide. The bis-conjugated ASO inhibits exon 7 splicing of the pre-mRNA SMN2 gene, which is responsible for the pathology of spinal muscular atrophy. These amphiphilic species exhibit a 4-fold increase in potency relative to the CPP-ASO mono-conjugate.

A similar approach is described by Wada *et al.*, who used cholesterol and GalNac as modifiers at the 5'-end of an ASO targeting ApoB (Figure 9).[129] The conjugation was prepared using phosphoramidite chemistry, amenable to solid support synthesis of oligonucleotides. The monovalent GalNAc phosphoramidite used was described earlier by Yamamoto *et al.*[130] Such a dual conjugation strategy reduces the renal biodistribution of the oligonucleotide. In that case, the cholesterol-conjugated ASO also possessed a 3-nucleotide-long linker with phosphodiester linkages susceptible to endonuclease hydrolysis in the targeted tissues or cells. Accumulation of the GalNac-modified ASO is low in the kidney just like the double 5'-modified cholesterol-GalNAc ASO. Renal accumulation is five times lower when the ASO is modified at both extremities one with the lipid and the other with GalNAc. As expected, the Gal-

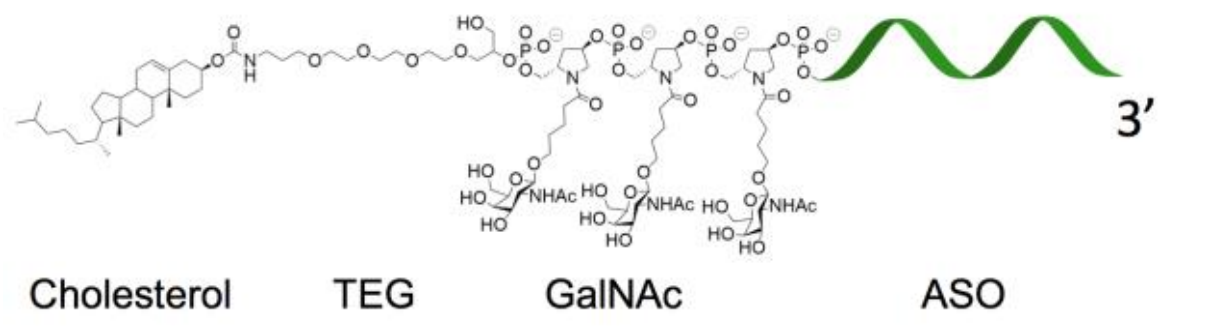

Figure 9: Cholesterol and GalNAc coupled ASO, adapted from Wada F. *et al.* 2018 [129] TEG : Triethylene glycol ; GalNAc : *N*-acetylgalactosamine ; ASO : AntiSense Oligonucleotide

Nac modified ASOs preferentially accumulate in the liver. Thus, the cholesterol–GalNAc dual strategy represents an effective strategy for reducing the nephrotoxic potential of ASOs, while maintaining the gene silencing activity in the liver.

***Spherical Nucleic Acids.*** Spherical Nucleic Acids (SNAs) are a class of nucleic acid conjugates differentiated from other conjugates here through their three dimensional nanostructure[131]. First developed by Chad Mirkin, this structure is composed of oriented single or double stranded nucleic acids in a dense spherical array (Figure 10). First generation SNAs possessed a gold nanoparticle core with Au-S linkages to the nucleic acids, and since then have been expanded to contain other cores for specific applications[131]. SNA structures are amenable to further modification to contain targeting ligands such as antibodies[132], and immune agents such as antigens[133,134]. SNAs can also be formed with RNA[135], modified backbones[133], and sugar modifications[136]. SNAs exhibit increased nuclease resistance due to their structure[131] and are readily endocytosed[137] into cells without additional transfection making them an attractive bioconjugate class to enhance delivery[137]. Liposomal SNAs[138] are formulated from 1,2-dioleoyl-sn-glycero-3-phosphocholine and a tocopherol-oligonucleotide conjugate which contains a phosphodiester linkage and a PEG spacer (discussed in its own section above). Tocopherol-oligonucleotide conjugates form micellar structures, but liposomal SNAs provide additional benefit[133]. Liposomal SNA formulations are being evaluated as a novel therapy for immune modulation[116,133,134] For example, SNAs are effective for both immune TLR agonism and inhibition through either target sequence dependent binding to TLR receptors or the receptor for mRNA translation[133].

Receptor agonism strategies focus on agonism through TLRs,[133,134] and the activity of the SNA conjugates is dependent on linker chemistry. SNA and tocopherol oligonucleotide conjugates show effective immune stimulation and tu-mor growth reduction *in vivo*,[133] and SNA formulations demonstrate the highest efficacy. In a recent publication[134], a melanoma-specific peptide antigen was conjugated via three different linkers to a TLR9 agonizing oligonucleotide adjuvant. The complimentary oligonucleotide was conjugated to an antigen.

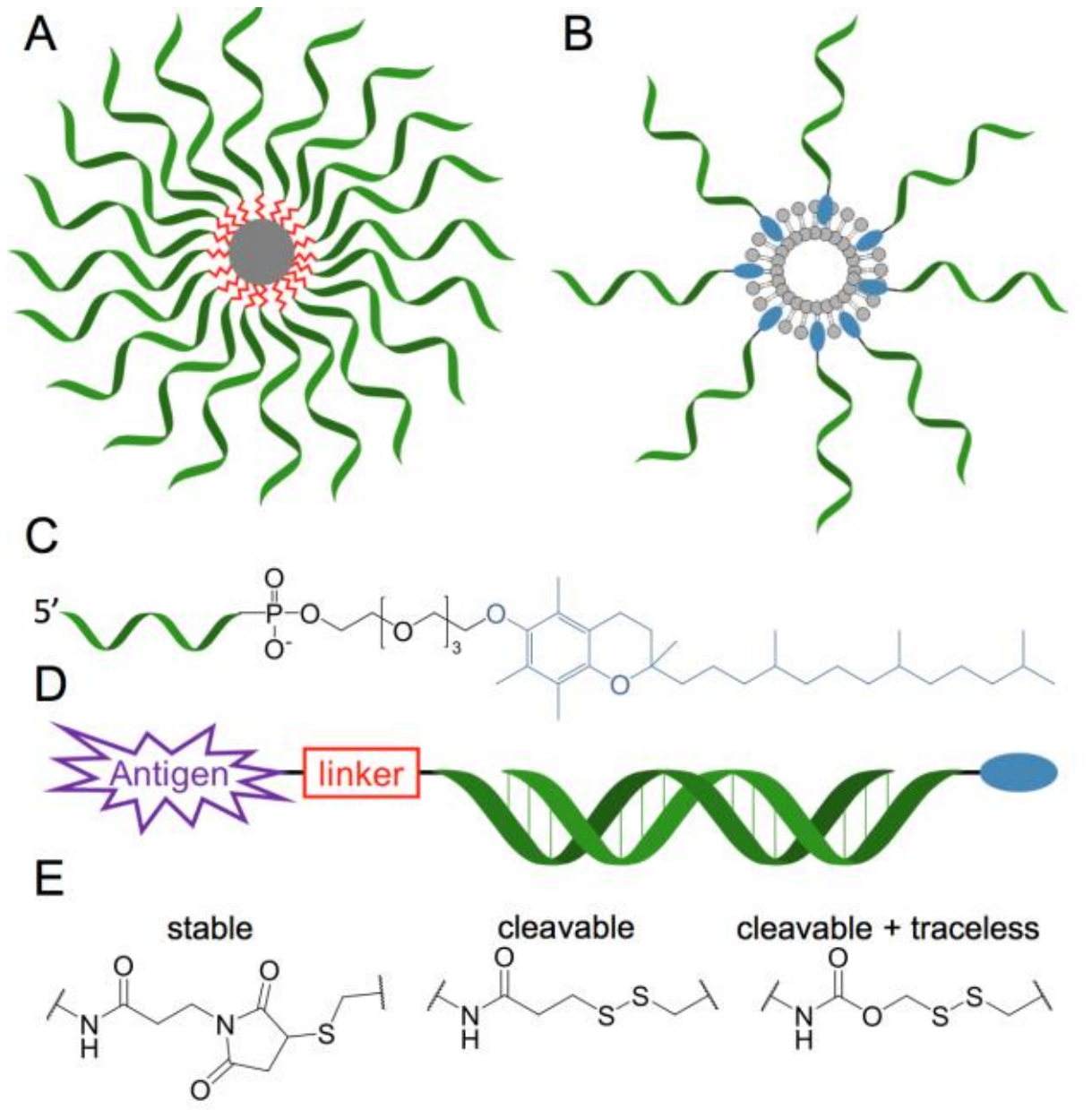

Figure 10: Spherical Nucleic Acids. A. SNAs present nucleic acids in an outward facing spherical array. B. Oligonucleotide-tocopherol conjugates form SNAs with liposomal cores. C. Structure of linkage and tocopherol moiety. D. TLR agonizing oligonucleotides conjugated to tocopherol can be combined with antigens through a complimentary oligonucleotide and chemical conjugation through a linker. E. Different conjugate linkage chemistries explored.

Three different antigen-oligonucleotide linkages were evaluated: one stable (N-(β-maleimidopropyloxy) succinimide ester), one cleavable (succinimidyl 3-(2-pyridyldithio)propionate), and a traceless linkage (4-nitrophenyl 2-(2-pyridyldithio)ethyl carbonate). See Figure 10 for structural comparison. The traceless linkage undergoes an internal cyclization in addition to a disulfide cleavage to yield the antigen's native N-terminus. The choice of linker conjugation significantly influenced activity. The SNA formulations possessing the cleavable, traceless delivery linkage yielded the greatest T cell activation and proliferation. Importantly, these findings highlight the critical role the linker plays. Moreover, they also demonstrate the potential for oligonucleotide conjugates to expand to new applications and a few SNA formulations are progressing through preclinical into early clinical development (e.g., NCT03086278, Table 1 and discussion below). [139,140].

## OLIGONUCLEOTIDES THERAPEUTIC APPLICATION

As exemplified in the preceding sections, numerous oligonucleotides have been investigated for a range of diseases. The majority of these bioconjugated oligonucleotides are in preclinical studies (i.e., *in vitro* and small animal *in vivo* efficacy studies). Importantly, several formulations are regulatory approved and used in the clinic (Table 1). The data accumulated in Table 1 are from company web sites and the clinicaltrials.gov web resource. The first regulatory approved oligonucleotide (in 1998 by the FDA and in 1999 by the EMA (*European Medicines Agency*)) was for the treatment of cytomegalovirus retina, a viral infection of the eye. If untreated, vision loss can occur in immune compromised patients. Specifically, Fomivirsen (Vitravene®) was developed by IONIS Pharmaceutical (ISIS at the time) in collaboration with Novartis Ophthalmics.[141] This biomacromolecule, composed of 21 PTO nucleotides, is complementary to one viral mRNA. The posology consists of a weekly intravitreal injection of 165 µg of a Fomivirsen solution at 6.6 mg/mL.[142] The market authorization was withdrawn on July 30th 2002 at the request of the manufacturer for business reasons. Vitravene® is still authorized in Switzerland and can be administered in the European Union upon specific request. Importantly, this success laid the foundation for a number of oligonucleotides being ushered into commercial development and the reader is referred to several reviews on this topic.[68,143–146]

As discussed in the introduction of this Review, the design of oligonucleotides for use in the clinic requires specific design considerations and optimizations other than improved delivery.[147] For enhanced *in vivo* efficacy, oligonucleotides containing chemical modifications such as 2'-O-MOE, 2'OMe, LNA and PMO are used to protect against degradation and increase target binding affinity. To date, the most common oligonucleotide modifications are PTO DNA and 2'-O-MOE. Additionally, 2'OMe and LNA are used, but most of these oligonucleotides are still in phase 1 or 2 clinical trials.[143,148] As of 2010, all of the oligonucleotides approved by the FDA and in phase 3 clinical trials contain PTO linkages.

The selection of oligonucleotide modifications defines the mechanism(s) of inhibition (Figure 2). The majority of oligonucleotides in phase 1 or 2 trial participate in an RNase H based inhibition mechanism, but are of more diverse chemical composition and functionalization. That said, the inhibition mechanisms available include intron-exon splicing, RNase H activation, or blockage of translation through steric hindrance. The following subsections are organized according to targeted disease with a discussion of the bioconjugated oligonucleotides used as well as their mechanism of action.

***Bone Marrow disorders.*** (Myelofibrosis and myelodysplastic disorders). Myeleofibrosis is caused by the replacement of bone marrow with scar tissue. The resulting scarring leads to a reduction of red blood cell formation. Myelodysplastic syndrome is a type of cancer in the bone marrow. Risk factors are genetic or caused by or arise during some cancer treatments. Imetelstat is a palmitoyl lipid oligonucleotide conjugate in clinic trials to treat these diseases. This 13 nucleotide-long ASO targets telomerase activity, using palmitate as the targeting ligand, with thio-phospharamidite linkages including the linkage to the palmitate group (Figure 11).

The drug has completed multiple phase II clinical trials for metastatic breast cancer and multiple myeloma cancers patients. The dose is limited by thrombocytopenia. Imetelstat (GRN163L) received a fast track status in October 2017 from the FDA for certain patients with a myelodysplastic syndrome.[149]

***Hepatitis B.*** Hepatitis B is a common liver infection. The genome of the hepatitis B virus (HBV) is an attractive target because there are multiple sites on the HBV genome that can be targeted simultaneously. There are several conjugates in the pipeline for the treatment of this disease. For example, ARC-520 is an equimolar combination of two cholesterol conjugated siRNAs, siHBV74 and siHBV77, targeting an 18 nucle-otide long sequence in the HBV genome. In addition to the siRNAs, a GalNAc conjugated peptide is included as an excipient to aid in delivery and endosomal escape. A Phase trial I was completed in 2018, but the results have not been publicized as of September 2018. Other clinical trials with ARC-520 have been withdrawn/terminated, likely due to the difficulty in knocking down expression due to HBV DNA integrated into the host genome. Arrowhead Pharmaceuticals has a combined phase I/II clinical trial currently recruiting for ARO-HBV. Alnylam terminated a phase I clinical trial using ALN-HBV01, a 23 nucleotide long, GalNAc conjugated, siRNA. This GalNAc conjugated siRNA incorporated numerous modifications for solubility and stability, including 2'-F, 2-OMe, with PS linkages only near the siRNA termini. The discontinuation is part of a new agreement with Vir Biotechnology to begin development on ALN-HBV02, also a GalNAc conjugate, citing reduced off target effects due to the incorporation of glycol nucleic acids (GNAs) in the backbone.

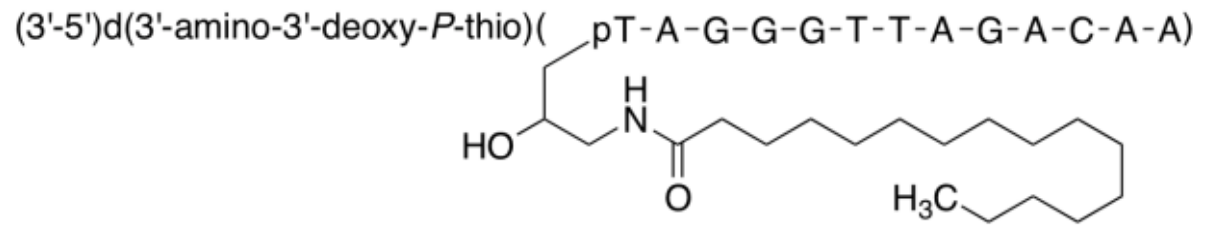

Figure 11: Structure of the thio-phospharamidite based ASO with a telomerase activity. The targeting moiety (palmitate) has been inserted at the 3’ extremity.

***Hemophilia.*** Hemophilia is a blood clotting disorder. Fitusiran (ALN-AT3sc) is a GalNAc conjugate targeting antithrombin currently entering phase III clinical trials. In a phase I study, the mean antithrombin (AT) knockdown was reported to be 59%. The study included 4 healthy volunteers (A) and 12 patients with severe hemophilia (B). In one hemophilia patient, a maximal knockdown of 86% was reported along with 114 days bleed free. This conjugate was administrated *via* subcutaneous injection. First phase results showed the reduction of levels of the antithrombin protein and the restoration of the balance for the production of clotting factors in adult A and hemophilia B patients, after a single monthly injection. The second phase of the trial was suspended because one hemophilia A patient died due to a cerebral edema. The suspension of the clinical trial was lifted after the trial’s protocol was amended to better mitigate risks. Alnylam and Sanofi are scheduled to conduct a phase 3 clinical trial in 2018/2019 with the ATLAS program,[150] which will include patients with hemophilia A and B with or without inhibitors, and patients previously receiving on-demand or prophylactic therapy.

**Psoriasis.** Psoriasis is a chronic skin disease where skin cells undergo a rapid cell cycle. SNAs are being developed to treat dermatologic indications such as mild to moderate Psoriasis[139,140]. Recently completed results from Phase I studies for IL17RA and TNFα protein targets are promising (Purdue Pharma). The SNAs are absorbed through the skin and deliver siRNAs into the epidermis[139] providing effective knockdown of targets within keratinocytes with minimal immune response. Topical administration offers a local treatment option, which minimizes systemic off-target effects.

**Cancer.** Cancer is the unwanted abnormal growth and invasion of cells in the body. Of the various treatment classes – surgery, radiation, chemotherapy, and immunotherapy – immunotherapy is a particularly exciting one as it provides a means to use the immune system to fight the disease. For example, activation of Toll-like receptor 9 (TLR9) initiates a cascade of innate and adaptive immune responses including increased interferon gene expression. Increased interferon gene expression correlates with improved responses to programmed death 1 (PD-1) inhibition. Thus, patients who did not respond or stopped responding to anti–PD-1 therapy are potential candidates for combined TLR9 agonists and anti–PD-1 therapy. Exicure has successfully completed a Phase I trial of TLR9 agonizing SNAs administered subcutaneously in healthy volunteers (NCT03086278). A Phase Ib/II study is planned to study intratumorally delivered SNAs both alone and in combination with intravenously administered pembrolizumab, an antibody targeting the PD-1 receptor. The target cohort is in patients that have not yet responded to anti-PD-1 or anti-PD-L1 therapies (NCT03684785).

**Table 1.** Bioconjugate oligonucleotides in the clinic.

| Name | Target | Linker | Moiety | stage | Trial Reference |
|---|---|---|---|---|---|
| APOC-III-L-Rx | apoC-III | | GalNAc3 | ii recriuting | NCT03385239 |
| ARC-520 | HBV | | Cholesterol | multiple i/ii | NCT01872065 |
| DCR-PHXC-101 | GO | | GalNAc | i recruiting | NCT03392896 |
| FXI-LRx | Factor XI | 2'-MOE | GalNAc | i recruiting | NCT03582462 |
| Inclisiran (ALN-PCSSC) | PCSK9 | | GalNAc | multiple i/ii/iii | NCT03397121 |
| Macugen | VEGF | | PEG | Approved | |
| Zimura (Avacincaptad pegol) | C5 | | PEG | iib | NCT03374670 |
| Fovista (E10030) | PGDF | | PEG | terminated | NCT01944839 |
| Imetelstat Sodium (GRN163L) | telomerase | thio-phosphoramidite | Palmitate | ii | NCT01731951 |
| Revusiran (ALN-TTRSC) | TTR-FAC | | GalNAc | discontinued | |
| Fitusiran (ANL-AT3sc) | antithrombin | | GalNAc | iii recruiting | NCT03549871 |
| Cemdisiran (ALN-CC5) | Complement C5 | | GalNAc | ii recruiting | NCT03303313 |
| Givosiran (ALN-AS1) | ALAS1 | | GalNAc | iii active | NCT03338816 |
| Lumasiran (ALN-G01) | GO | | GalNAc | ii enrolling | NCT03350451 |
| ALN-HBV | HBV | | GalNAc | i terminated | NCT02826018 |
| ALN-AAT | alpha-1antitrypsin PiZZ allele | | GalNAC | i/ii terminated | NCT02503683 |
| RG-101 | HCV | | GalNAc | terminated | |
| ALN-TTRSC02 | TTR | | GalNAc | i completed | NCT02797847 |
| Akcea-APO(a)-L_rx | apoC-III | 2'-O-(2-methoxyethyl) | GalNAc | ii active | NCT03070782 |
| IONIS-AGT-LRx | AGT | 5-10-5 MOE gapmer | GalNAc | i completed | NCT03101878 |
| IONIS-ANGPTL3-LRx | ANGPTL3 | Phosphtiorate and Phos-phodiester | GalNAc | ii | NCT03371355 |
| IONIS-TMPRSS6-LRx | TMPRSS6 | | GalNAc | i | NCT03165864 |
| ARO-HBV | HBV | | | i/ii, recruiting | NCT03365947 |
| IONIS-PKK-LRx | PKK | | | i recruiting | NCT03263507 |
| IONIS-GHR-LRx | GHR | | | ii recruiting | NCT03548415 |
| IONIS-AZ4-2.5-LRx | undisclosed | | | i recruiting | NCT03593785 |
| AMG-890 | cardiovascular dis-ease | | | i recruiting | NCT03626662 |
| ARC-AAT | AAT | | Cholesterol | ii, withdrawn | NCT02363946 |
| AST 008 | TLR9 | | Spherical Nucleic Acids | i completed | NCT03086278 |
| AST 008 and Pembroli-zumab i | TLR9 | | Spherical Nucleic Acids | i planned | NCT03684785 |
| XCUR17 | IL17RA and TNFα | | Spherical Nucleic Acids | i completed | |

***Age-Related Macular Degeneration.*** Age-related Macular Degeneration (ARMD) is the gradual worsening of vision and eye health over time due to retina damage. The blurring or loss of vision associated with ARMD is characterized by the proliferation of blood vessels behind the retina. In "wet" macular degeneration, neovascularization can worsen symptoms and is activated by Vascular Endothelial Growth Factor (VEGF) signaling. Macugen® (Pegaptanib) is a 27-nucleobase aptamer targeting $VEGF_{165}$ binding,[151,152] and possesses a 3'-3' deoxythymidine extremity and a 5'end with two branched polyethylene glycol chains of 20 kDa. The 2' positions are modified for nuclease resistance purposes by 2'OMe for bases purines and fluorinated pyrimidines (Figure 12). Importantly, the aptamer binds the heparin site of VEGF with a picomolar affinity. The FDA approved this aptamer for the treatment of "wet" or Age-Related Macular Degeneration. Specifically, Macugen®, injected intravitreally, locally reduces the proliferation/*expansion* of blood vessels and a lack of adverse events is still observed at a dose ten times higher than clinically used. The product is available in the form of pre filled syringes of 90 µL solution containing 0.3 mg of active substance.

Several other drugs have been developed for the treatment of ARMD. In the EU, Lucentis® (ranibizumab, Novartis), a monoclonal antibody fragment that targets VEGF, and Eylea® (aflibercept, Bayer) a recombinant fusion protein that binds to VEGF-A, VEGF-B, and placental growth factor (PIGF) with higher affinity than their native receptors. Avastin® (bevacizumab, Roche) is a humanized recombinant monoclonal antibody that binds VEGF and competitively inhibits the binding between VEGF and Flt-1 and KDR. It adds to the list of Macugen® competitors in the United States. The other nucleic acid treatment for ARMD is a modified siRNA called bevasiranib. It is a duplex of a 21-nt long RNA that targets VEGF.[153] Bevasiranib has not been approved but shows efficacy when combined with Lucentis® in clinical trials.[148] The effects of bevasiranib appear only six weeks after the beginning of the treatment, but the combination of Lucentis®/bevasiranib provides better visual acuity than Lucentis® alone.

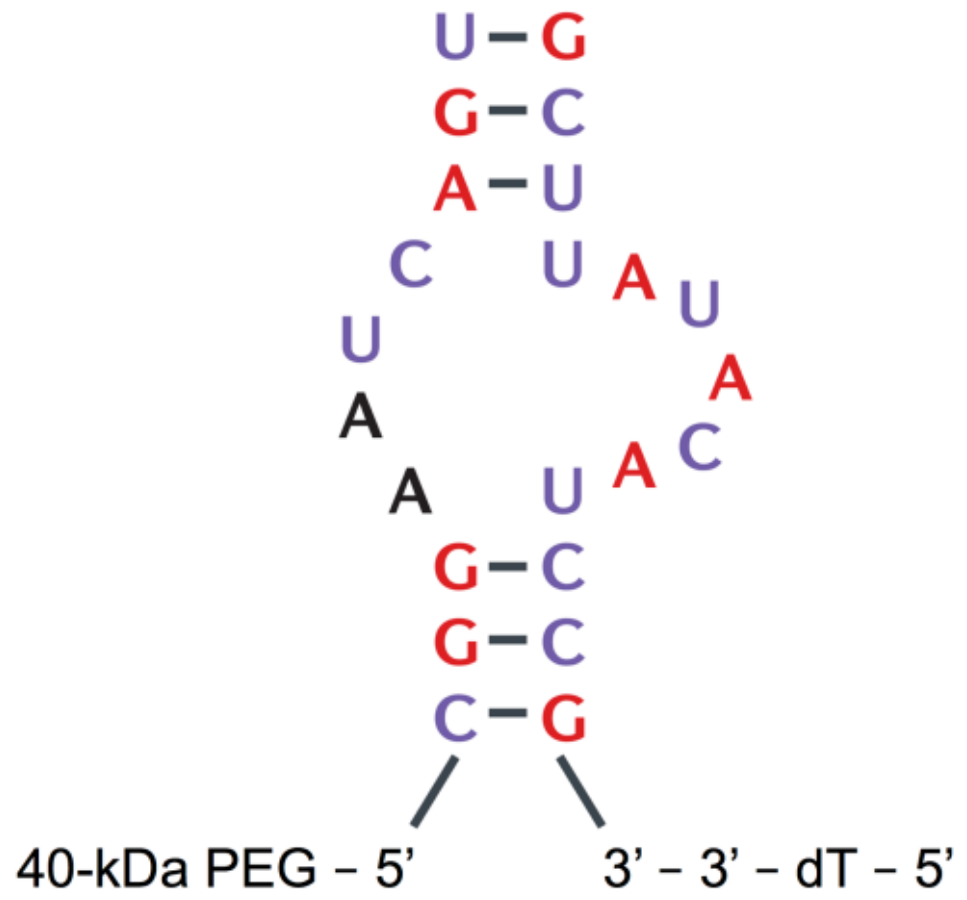

Figure 12: Secondary structure of Pegaptanib, active substance of Macugen® Adapted from Ng E. W. M. *et al.* 2006 [152]

***Hepatic Porphyrias.*** These conditions are a combination of rare diseases that all overlap in that they all feature the accumulation of porphyrins in the blood. The build-up of these proteins causes nervous or dermatological symptoms. Currently in an ongoing phase III clinical trial, Givosiran (ALN-AS1), is a GalNAc conjugate siRNA targeting aminovulinate synthase 1 (ALAS1) present in the liver. In a Phase I trial with 23 patients, 11 patients reported mild or moderate adverse events (AE) with 1 reporting an unrelated, severe AE of abdominal pain and another reported unrelated AE of bursitis. Common mild/moderate AEs of abdominal pain, diarrhea, nasopharyngitis, and hypoesthesia, pruritus, and rash were noted, but were not serious enough to stop treatment. The product was injected by single or multiple subcutaneous injections and compared to an injection of sterile normal saline solution. The purpose of the next phase 1 trial was to evaluate the safety and tolerability of Givosiran (ALN-AS1) in Acute Intermittent Porphyria (AIP) patients as well as to characterize pharmacokinetics (PK) and pharmacodynamics (PD) of ALN-AS1 in AIP patients. The safety and efficacy of Givosiran was evaluated in 17 patients with acute AIP. The patients were initially monitored for three months and those who experienced one porphyria attack were eligible to receive once monthly injections of 2.5 or 5.0 mg/kg Givosiran or placebo for six months. Durable inhibition of ALAS1 was observed after Givosiran treatment. The lower tested dose of 2.5 mg/kg reduced the annualized attack rate by 83% and afforded a reduction of 88% of hemin compared to the placebo. A larger dose did not produce any amelioration. In 2018, data showed that at 22 months of therapy, patients were still experiencing a robust beneficial effect from Givosiran.

***Famyloid polyneuropathy***. Familial Amyloid Polyneuropathy (FAP) is a genetic autosomal dominant disorder resulting from a mutation of the TTR (transthyretin) gene, leading to TTR protein aggregation, which negatively affects the nervous system. Oligonucleotide therapeutic agents designed for the treatment of FAP, also known as Transthyretin-related hereditary amyloidosis (ATTR), target the mRNA sequence of transthyretin (TTR) for degradation. Three formulations are in clinical trials to decrease expression of TTR: Inotersen (developed by Ionis and GSK), Onpattro (Patisiran developed by Alnylam), and Revusiran (developed in collaboration between Alnylam and Sanofi Genzyme).[154] These therapeutics (2 siRNAs -one being a GalNac conjugate- and an ASO) are in phase 3 clinical trials. One study compared the inhibitory activity of Inotersen (ASO) and Patisiran (siRNA). Both treatments led to a 80% decrease in TTR expression.[155] Inotersen (IONIS-$TTR_{Rx}$),[156] is a 20-nucleotide long gapmer PTO ASO with 5 2'-O-MOE modifications at both extremities. The ASO targets pre-mRNA that is degraded following RNase H activity.[157] Inotersen sodium is subcutaneously injected weekly at a dose of 300 mg (284 mg of inotersen). Of note, three cases (out of the 172 patients tested over 64 weeks) reported thrombocytopenia. Inotersen has received marketing authorization approval from the European Commission (EC) for the treatment of stage 1 or stage 2 polyneuropathy in adult patients with hereditary transthyretin amyloidosis (hATTR). Inotersen is commercialized under the name of Tegsedi™.

Patisiran (ALN-TTR02, Onpattro™) is an advanced siRNA under development and is administered intravenously as lipidic nanoparticles (LNP) at noticeably low doses of 0.3 mg/kg every three weeks over an 18-month period. The most common adverse events observed in patients were peripheral edema and infusion related reactions.[148,158] In August 2018, Patisiran received U.S. Food and Drug (FDA) approval for the treatment of the polyneuropathy of hereditary transthyretin-mediated amyloidosis in adults. Regulatory filings in other markets, including Japan, are planned. The European Commission also granted marketing authorization for Onpattro™ on August 30th, 2018.

Revusiran. a GalNac siRNA conjugate administered subcutaneously, was previously undergoing phase 3 clinical trials. This molecule has two PTO linkages, 22 2'-OMe nucleotides, and another 22 modified residues with 2'F modifications. However, development was stopped due to numerous AEs leading to deaths in phase 3 clinical studies. Causes of theses side effects have not been published but the high weekly dose of 500 mg was suspected.

**Historic timeline of developments**

There have been numerous developments over the past 50 years that have led to the clinical efficacy of oligonucleotides for treating diseases.[159] The timeline featuring significant accomplishments is presented in figure 12. ASOs were the first oligonucleotides to reach clinical trials and to receive regulatory approval. Notably, this conjugated (pegylated) aptamer has been in use for more than 14 years. The development of siRNA therapeutics advanced significantly in the early 2000s and their first use in humans occurred in 2005 (bevasiranib). siRNA therapeutics that functioned as splicing modulators followed next in development and commercialization. Antagomirs (miRNA inhibitors) were developed in 2009 and were followed by miRNA mimetics in 2013. These two former classes of oligonucleotides utilize miRNAs to modulate the activity of a given gene. For example, Miravirsen sequesters microRNA-122 and is used for the treatment of chronic hepatitis C,[160,161] whereas mimics of microARN-34a exhibit potentiated antitumor activity. Over 30 oncogenes are known to be inhibited by this microRNA.[162]

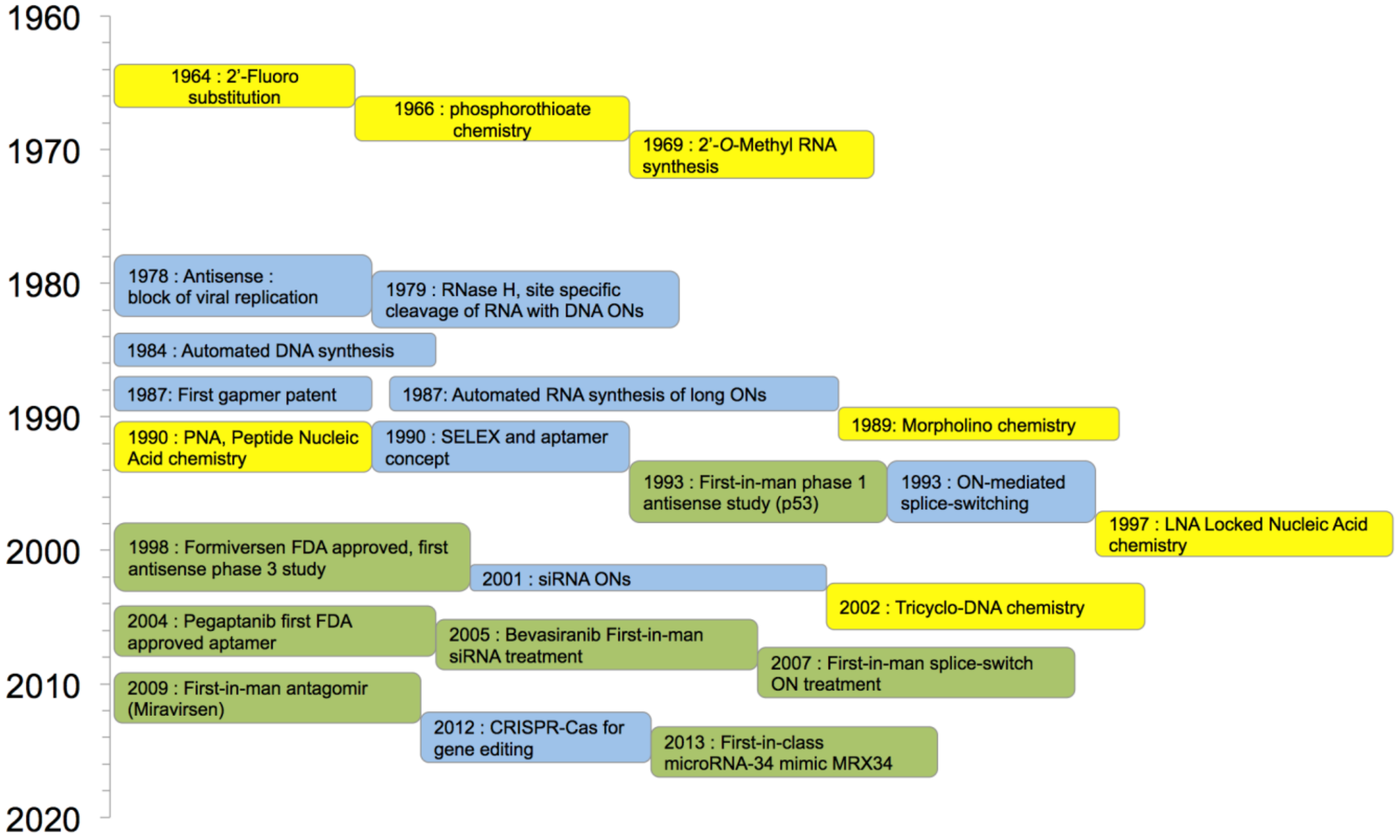

Figure 13: Timeline for Key Advances in Therapeutic Oligonucleotides. Adapted from Lundin K. E. *et al.* 2015 [159] The various substitutions and the advances of chemical modifications are shown in yellow, main discoveries and technological achievements are noted in blue and clinical studies are highlighted in green.

## 6. FUTURE OPPORTUNITIES AND CONCLUSION

Several bioconjugated oligonucleotides are regulatory approved and used in the clinic. These bioconjugates possess a carbohydrate (Patisiran), a polymer (Pegaptanib), or a lipid (Imetelstat - GRN163L, currently on fast track status) covalently linked to the oligonucleotides. A number of additional ones are in preclinical development as well as clinical studies (see Table 1.). The primary function of these added conjugates is to enhance targeting, facilitate vectorization, reduce oligonucleotide susceptibility to nuclease activity, eliminate or reduce need for toxic transfection agents, and generally improve pharmacological activity and efficacy. Increasing the therapeutic window and reducing the likelihood for adverse events is a top priority, and the use of conjugated oligonucleotides is an effective approach. As discussed above, these conjugates are lipophilic or a targeting moiety.

Most of the therapies currently in late-stage development target the liver and liver associated diseases. This is largely due to the success of the GalNAc modification, and the fact that the liver is an efficient filter for the blood. Other highly efficient targeting moieties need to be developed to target other organs or tissues – lung, kidney, heart, bone marrow, brain, muscle, etc. Poor delivery to target sites necessitates higher doses, decreasing safety with greater chances of AEs. One strategy to reduce the severity of AEs is to modulate or knockdown the effect after administration. For example, a single subcutaneous 3 mg/kg GalNAc-siRNA dose can be effectively reversed with a 0.1 mg/kg GalNAc oligonucleotide conjugate targeting the seed region of the incorporated siRNA strand[163]. The result demonstrates both effective knockdown and return of TTR protein levels in preclinical *in vivo* studies. The evolution of safe antidotes for oligonucleotide therapies, such as the REVERSIR strategy reported by Zlatev *et al.*, poses benefits for safety and adoption into the clinic.[164] The possibility to finely control RNAi pharmacology is a desired feature for highly efficient nucleic based therapeutic treatments.

At the molecular level, advances in oligonucleotide chemistry are providing new methods and site-specific reactions to modify oligonucleotides at the base, sugar, phosphate, or end groups. Additionally, natural bases can be substituted with synthetic analogs. These developments afford a large diversity of new and unique biomacromolecules. Still, significant opportunities exist in this area to improve on reaction efficiency, to expand on the compositional diversity of modifications, and to realize structure-property relationships for nanoscale, self-assembling, and supramolecular systems. These improvements will allow for more modular, diverse, and effective formulations. Similarly, molecular biology studies are providing new therapeutic targets with an improved understanding of pathways and resulting pathologies. As such, the development of combination therapies is forthcoming. The possible combinations will target either one or multiple mRNAs or combine a bioconjugated oligonucleotide with a small molecule or protein in order to improve efficacy and safety profiles.

As discussed in this review, modifications such as GalNAc, cell penetrating peptides, α-tocopherol, aptamers, antibodies, cholesterol, squalene, fatty acids, and nucleolipids are new promising therapeutic candidates in the field. Despite the clinical progress to date, the design of modified oligonucleotides encompassing both the optimal delivery and efficient biological activity remains an important challenge. Diseases such as coronary heart disease, lower respiratory infections, cancer, diabetes, and Alzheimer's are in need of new therapies, and, as oligonucleotides can target and bind nearly any expressed mRNA, the possibility to develop oligonucleotide therapeutics exist. Key to these future successes will be understanding the pharmacokinetic and biodistribution profiles of these new therapeutics. Despite the promising bioconjugated oligonucleotides reported so far, oligonucleotide delivery represents a major hurdle still to be addressed. As mentioned previously, most of the oligonucleotide conjugates used in clinic bio-accumulate in the liver. Furthermore, a clear relationship between the composition and structure of the chemical modification and its resulting effect on pharmacokinetic/biodistribution must be both quantified and established with appropriate positive and negative controls in place. For example, membrane interacting bioconjugates often interact with many targets and are known for their promiscuity. Such a molecular rationale would allow scientists to design the chemical structure of the oligonucleotides depending on the specific disease, target tissue/organ, cell type etc.

In summary, the use of oligonucleotides in the clinic is in its infancy and the potential for significant advances in patient care exist. Given the clinical success of monoclonical antibodies (mAb) as a biologic, we speculate that oligonucleotides will have a greater impact due to the specificity inherent to the target sequence, the ability to target nearly any protein, and their stability both pre-administration and *in vivo*. The full realization and maturation of the bioconjugated oligonucleotide therapeutic field will require time – e.g., 19 mAb drugs were on the market only after 20 years of development – but this should not curtail enthusiasm. In fact, findings from these mAb studies as well as from polymer conjugated enzymes and small molecules used in the clinic will catalyze development and subsequent clinical applicability.[165] Bioconjugation chemistry is at the centerpiece on this therapeutic oligonucleotide revolution. We encourage all to investigate this area where conceptualizing, designing, and engineering at the molecular level new functionalized oligonucleotides is key to success in the clinic.

## AUTHOR INFORMATION

### Corresponding Author

* Mark W. Grinstaff, mgrin@bu.edu
* Philippe Barthélémy, philippe.barthelemy@inserm.fr

## ACKNOWLEDGMENT

This work was supported in part by funding from the Inserm transfert (IT), the National Institute of Health T32 Grant entitled Translational Research in Biomaterials (NIH T32EB006359, AM),[166] and the Distinguished Professor of Translational Research Chair at Boston University (MWG). The authors (PB and SB) thank Dr Lara Moumne (IT) for fruitful discussions.

TOC

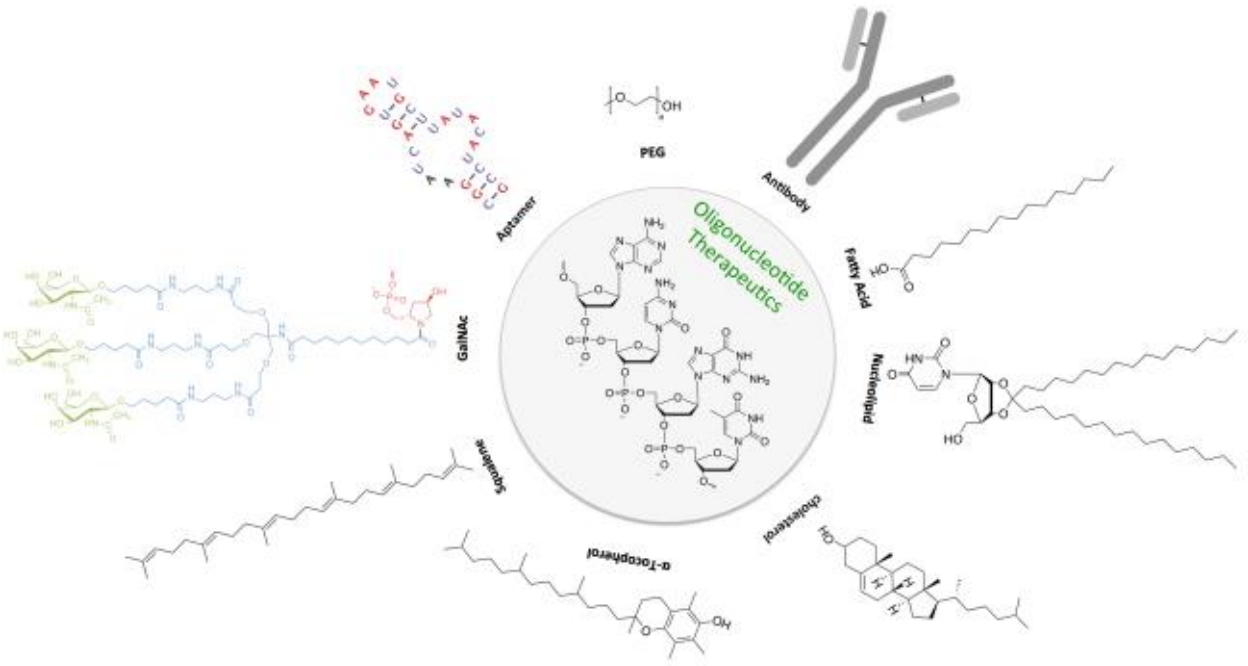